\documentclass[journal,onecolumn]{IEEEtran}
\usepackage{mathtools,amsthm,amssymb,amsfonts,bm}
\usepackage{amsfonts,amsmath,amssymb,amsthm,mathrsfs}
\usepackage{thmtools,amsmath,extarrows}
\usepackage{algorithmic}
\usepackage{algorithm}
\usepackage{array,tabularx,booktabs,multirow,colortbl}
\usepackage{makecell}
\usepackage{tikz-cd,physics}
\usepackage[all]{xy}
\usepackage[caption=false,font=normalsize,labelfont=sf,textfont=sf]{subfig}
\usepackage{hyperref}
\hypersetup{hidelinks,colorlinks=true,linkcolor=blue,citecolor=blue,urlcolor=blue}
\usepackage{textcomp}
\usepackage{stfloats}
\usepackage{url}
\usepackage{verbatim}
\usepackage{graphicx}
\usepackage[noadjust]{cite}
\makeatletter
\newtheoremstyle{mythm}{3pt}{3pt}{}{16pt}{\bfseries}{:}{.5em}{}
\theoremstyle{mythm}
\newtheorem{theorem}{Theorem}
\newtheorem{example}{Example}
\newtheorem{definition}{Definition}
\newtheorem{remark}{Remark}

\newtheorem{corollary}{Corollary}
\newtheorem{lemma}{Lemma}

\newtheorem{construction}{Construction}

\newcommand{\cA}{\mathcal{A}}

\newcommand{\cC}{\mathcal{C}}

\newcommand{\cF}{\mathcal{F}}

\newcommand{\cP}{\mathcal{P}}

\newcommand{\cU}{\mathcal{U}}
\newcommand{\cV}{\mathcal{V}}
\newcommand{\cW}{\mathcal{W}}

\newcommand{\F}{\mathbb{F}}

\renewcommand{\le}{\leqslant}
\renewcommand{\leq}{\leqslant}
\renewcommand{\ge}{\geqslant}
\renewcommand{\geq}{\geqslant}
\newcommand{\ceilenv}[1]{\left\lceil #1 \right\rceil}
\newcommand{\floorenv}[1]{\left\lfloor #1 \right\rfloor}

\newcommand{\parenv}[1]{\left( #1 \right)}

\newcommand{\opt}{\mathrm{opt}}

\newcommand{\unchanged}{\mathcal{U}}
\newcommand{\reading}{\mathcal{D}}
\newcommand{\written}{\mathcal{W}}

\begin{document}
\title{Locally Repairable Convertible Codes: Improved Lower Bound and General Construction}
\author{
Songping Ge, Han Cai, \textit{Member, IEEE}, and Xiaohu Tang, \textit{Fellow, IEEE}
\thanks{S. Ge is with the School of Mathematics, Southwest Jiaotong University, Chengdu, 610031, China (e-mail: gesongping@my.swjtu.edu.cn).}
\thanks{H. Cai and X. Tang are with the School of Information Science and Technology, Southwest Jiaotong University, Chengdu, 610031, China (e-mail: hancai@aliyun.com; xhutang@swjtu.edu.cn).}
}
\maketitle

\begin{abstract}
In this paper, we consider convertible codes with the locally repairable property. We present an improved lower bound on access cost associated with $(r,\delta\ge 2)$-locality, which extends the known lower bound only related to $r$. We then provide a general construction of convertible codes with optimal access cost which shows that these codes can have super-linear length or maximum repairable property. Furthermore, we propose explicit constructions of convertible codes with the final code achieving super-linear length or maximum repairable property when the initial codes have super-linear length or maximum repairable property respectively. Specially, compared with known constructions, our construction is applicable to the case $\delta>2$ for the first time.
\end{abstract}


\section{Introduction}

The proliferation of large-scale cloud storage systems and distributed file architectures (e.g., Windows Azure Storage, Google File System) has transformed disk failures from exceptional events into commonplace occurrences. To handle these failures, conventional replication strategies create multiple copies of data blocks across storage nodes, albeit at substantial storage costs. This limitation has driven the development of erasure coding techniques that achieve comparable reliability with significantly improved storage efficiency.

Considering various aspects of distributed storage system performance, researchers have introduced a range of coding techniques. For instance, to maximize failure tolerance while minimizing redundancy, they have explored maximum distance separable (MDS) codes. On the one hand, to reduce the data transmitted during the repair process, regenerating codes with optimal repair bandwidth have been proposed~\cite{dimakis2010network}. On the other hand, to limit the number of nodes involved in data recovery, locally repairable codes (LRCs) have been designed~\cite{gopalan2012locality}. Over the past decade, significant progress has been made in coding techniques for distributed storage systems based on these performance criteria. For further details, refer to the following: Codes with optimal repair bandwidth~\cite{dimakis2010network,guruswami2017repairing,tamo2018repair,ye2017explicit,tamo2017optimal,li2018generic,li2016optimal,tamo2012zigzag,
zeh2016bounds,liu2023generic,hu2017optimal,hou2019rack,chen2020explicit,wang2023rack,DBLP:journals/tit/LiWHY24};
Optimal locally repairable codes~\cite{huang2013pyramid,gopalan2012locality,rawat2013optimal,tamo2014family,cadambe2015bounds,kim2018locally,
li2019construction,xing2019construction,cai2020optimal,cai2021optimal_GPMDS,chen2020improved,hao2020bounds,sasidharan2015codes};
Codes with good availability~\cite{rawat2016locality,TaBaFr16bounds,cai2018optimal,cai2019optimal,jin2019construction}; Codes with optimal update and access properties~\cite{tamo2012zigzag,mazumdar2014update,li2015framework,chen2020enabling} and references therein.

All these coding schemes are designed under a fixed code rate, which corresponds to a preset failure rate for the storage system. However, device failure rates are inherently dynamic. Thus, we usually preset a low code rate corresponding to the worst device failure rate,
which may cause inefficiency during the device's stable period. To overcome this drawback, Kadekodi et al.~\cite{kadekodi2019cluster} introduced a cost-effective framework that can dynamically adapt code rates and redundancy levels to the fluctuations in the device failure rate, thereby optimizing storage efficiency. This process, termed \textit{code conversion} \cite{maturana2022convertible}, requires transitioning encoded data from an initial code $[n_I,k_I]$ to a final code $[n_F,k_F]$ with distinct parameters. Code conversion operates in two primary modes:
\begin{enumerate}
    \item \textit{Merge regime}: Merging multiple initial codewords into a single final codeword.
    \item \textit{Split regime}: Splitting an initial codeword into multiple final codewords.
\end{enumerate}

The traditional code conversion method involves re-encoding the original data from an initial code $\mathcal{C}_I$ to the final code $\mathcal{C}_F$. Nevertheless, this approach requires retrieving a vast number of symbols, decoding the entire dataset, transmitting the data across the network, and subsequently re-encoding it. This results in considerable overhead in terms of network bandwidth, I/O load, and CPU resource utilization within the cluster, hence impeding the system's efficiency. To address these limitations, Maturana and Rashmi~\cite{maturana2022convertible} pioneered the theory of \textit{convertible codes} enabling parameter transitions without full data reconstruction. Building upon their framework, the studies \cite{maturana2023bandwidth,maturana2023locally,maturana2020access} established fundamental theories for MDS code conversions. Subsequently, \cite{kong2024locally} and \cite{ge2024mds} provided a class of convertible codes with optimal read and write access costs based on RS codes and the extended generalized RS codes \cite{huffman2010fundamentals}, respectively.

Analogous to MDS codes, storage systems employing LRCs face the inefficiency of using a code with a fixed rate, but the failure rate of storage devices changes over time. Consequently, the design of LRCs that facilitate efficient code conversions is of both practical and theoretical significance. Xia et al. \cite{xia2015tale} first proposed a conversion method from the initial (2, 2)-locality to the final (6, 2)-locality while minimizing the number of node read. Subsequent constructions mainly considered the same locality between the initial and final codes. In \cite{maturana2023locally}, Maturana and Rashmi proposed a new construction technique for designing \textit{locally repairable convertible codes} (LRCCs) that can perform code conversion at a lower cost than the default approach by direct computation, projection, piggybacks, projected piggybacks, linear combination, and instance reassignment. Recently, Kong \cite{kong2024locally} established a lower bound on the access cost of LRCCs and proposed constructions of LRCCs that can perform code conversions with access cost achieving this bound with equality based on LRCs introduced in \cite{tamo2014family}.

The long-standing MDS conjecture asserts that no non-trivial MDS codes (with distance $d > 2$) exist beyond length $q+1$ over an alphabet of size $q$, with precisely two exceptional cases: when $q$ is even with $k = 3$, or $k = q-1$, the maximal allowable length extends to $q+2$. This conjecture was resolved for prime fields by Ball \cite{ball2012Large}. For MDS convertible codes, Ge et al., \cite{ge2024mds} presented constructions with field size $q = n_F - 1$, where $n_F$ is the code length of the final code. The situation differs fundamentally for optimal LRCs. The studies \cite{guruswami2019How,cai2020optimal,kong2021new} proposed explicit constructions of optimal LRCs with super-linear length, i.e., $n=O(q)$. This naturally raises the first open question \cite{kong2024locally}:

\textbf{Question 1: Can locally repairable convertible codes satisfy that the final code preserves super-linear length when the initial codes have super-linear length?}

While the benefits of locality are well-established, this feature entails a critical trade-off between the code rates and minimum Hamming distances. The \textit{maximally recoverable} (MR) codes \cite{gopalan2012locality} can correct erasures beyond their guaranteed minimum distance. As a specialized subclass of LRCs, MR codes can correct all erasure patterns that are information-theoretically correctible given the prescribed dependency relations between data symbols and parity symbols. This naturally raises the second question:

\textbf{Question 2: Can locally repairable convertible codes satisfy that the final code preserves maximally recoverable property when the initial codes inherently exhibit maximally recoverable property?}

To address these two questions, we first establish an improved lower bound on the access cost for convertible codes with \((r,\delta)\)-locality. A comprehensive comparison with existing bounds is presented in Table \ref{compare bounds}. We then propose the existence conditions and a general construction for access-optimal locally repairable convertible codes with a special class of parameters. Finally, as concrete examples, we construct explicit convertible codes with super-linear length (i.e., $n_F=\Omega(q^\delta)$ and $n_I=\Omega_t(n_F/t)$, where $n_F$ and $n_I$ are the code lengths of the final code and initial codes respectively, and $t$ is the number of initial merged codewords), or maximally recoverable property (i.e., both the final and initial codes are MR codes), based on the general construction. To the best of our knowledge, these are the first explicit constructions of LRCCs with super-linear length or maximally recoverable property. In Table \ref{compare construction}, we summarize and compare known constructions with ours which are the first constructions with $\delta > 2$.

The remainder sections of the paper are organized as follows: Section \ref{sec pre} reviews notation and preliminaries. Section \ref{sec bound} presents improved lower bounds on access cost for locally repairable convertible codes in the merge regime. Section \ref{sec gen con} proposes a general construction of access-optimal locally repairable convertible codes. Section \ref{sec exp con} proposes explicit access-optimal locally repairable convertible codes with super-linear length or maximally repairable property.
Finally, Section \ref{conclusion} concludes this paper.

\renewcommand{\arraystretch}{1.5}
\begin{table}[t!]
    \centering
    \caption{Lower Bounds on The Access Cost for Convertible Codes within Merge Regime ($k_F=tk_I$)}
    \begin{tabular}{|c|c|c|c|}
    \hline
    \textbf{Code description} & \textbf{Read access cost $\rho_r$}  & \textbf{Write access cost $\rho_w$} & \textbf{Reference} \\
    \hline
    \hline
    \multirow{3}{*}{\makecell{$\mathcal{C}_I$: $[n_I,k_I]_q$ code \\ $\mathcal{C}_F$: $[n_F,k_F,d]_q$-LRC \\ with $(r,\delta)_a$-locality}} & \multirow{3}{*}{\makecell{ $\rho_r\ge \begin{cases}k_F,  \mbox{if } \overline{\Delta}\le 0 \text{ or } d>n_I-k_I+1, \\ t\left(k_I-\lceil\frac{r\overline{\Delta}}{r+1}\rceil\right), \mbox{otherwise}, \end{cases}$ \\ where $\overline{\Delta}=n_F-2d-(t-1)k_I+3-\lceil\frac{(t-1)k_I}{r}\rceil$} } &  \multirow{3}{*}{$\rho_w\ge n_F-t(\overline{\Delta}+d-1)$} & \multirow{3}{*}{\cite[Theorem IV.4]{kong2024locally}}  \\
    & & &\\
    & & &\\
    \hline
    \multirow{3}{*}{\makecell{$\mathcal{C}_I$: $[n_I,k_I]_q$ code \\ $\mathcal{C}_F$: $[n_F,k_F,d]_q$-LRC \\ with $(r,\delta)_a$-locality}} & \multirow{3}{*}{\makecell{ $\rho_r\ge \begin{cases}k_F,  \mbox{if } \Delta\le 0 \text{ or } d>n_I-k_I+1,\\ t\left(k_I-\Delta + (\delta-1) \left\lfloor \frac{\Delta}{r+\delta-1} \right\rfloor\right), \mbox{otherwise}, \end{cases}$ \\ where $\Delta=n_F-2d-(t-1)k_I+2-(\lceil\frac{(t-1)k_I}{r}\rceil-1)(\delta-1)$} } &  \multirow{3}{*}{$\rho_w\ge n_F-t(\Delta+d-1)$} & \multirow{3}{*}{Corollary \ref{thm4}}  \\
    & & &\\
    & & &\\
    \hline
    \end{tabular}
    \label{compare bounds}
\end{table}

\renewcommand{\arraystretch}{1.5}
\begin{table}[t!]
    \centering
    \caption{Known Constructions of $(t,1)_q$ Locally Repairable Convertible Codes within Merge Regime with Parameters $(n_I,k_I,r_I,\delta_I;n_F,k_F=tk_I,r_F,\delta_F)_a$}
    \begin{tabular}{|c|c|c|c|}
    \hline
    \textbf{Construction} & \textbf{Restrictions}  & \textbf{Field size $q$ requirement} & \textbf{Additional properties}\\
    \hline
    \hline
    \multirow{3}{*}{\makecell{ Access optimal LRCCs \\ \cite[Construction III,IV]{kong2024locally} }} & \multirow{3}{*}{\makecell{ $r_I=r_F=r$ \\ $\delta_I=\delta_F=2$ \\ $r|k_I, r|k_F$ \\ $(r+1)|n_I, (r+1)|n_F$ }} &  \multirow{3}{*}{\makecell{ $\frac{k_I(r+1)}{r} ~|~ (q-1)$ and \\ $q\ge \frac{k_I(r+1)}{r} \max\{ t+1, \lceil\frac{rn_I}{(r+1)k_I}\rceil+1 \}+1$ }} &  \\
    & & & \\
    & & & \\
    \hline
    \multirow{5}{*}{\makecell{ General access optimal LRCCs \\ (Construction \ref{con})}} & \multirow{5}{*}{\makecell{ $t|\ell, \ell/t>g-\ell\ge h\ge 0$ \\ $r_I=r_F=r$ \\ $\delta_I=\delta_F=\delta\ge 2$ \\ $n_I=(\ell/t+g-\ell)(r+\delta-1)$ \\ $n_F=(\ell+h)(r+\delta-1)$ \\ $k_I=\ell r/t, k_F=\ell r$ }} &  \multirow{5}{*}{\makecell{ Determined by the field size \\ requirement of $\mathcal{C}_I$ and $\mathcal{C}_F$ }} & \multirow{5}{*}{\makecell{ super-linear length \\ (Theorem \ref{cor5}) \\ or maximally recoverable \\ (Theorem \ref{cor3}) }} \\
    & & & \\
    & & & \\
    & & & \\
    & & & \\
    \hline

    \multirow{4}{*}{\makecell{ LRCCs with information locality\\ \cite{xia2015tale} }} & \multirow{4}{*}{\makecell{ $t=1$,\\ $r_I=2,r_F=6$\\ $\delta_I=\delta_F=2$\\ $n_I=20,n_F=16$\\ $k_I=k_F=12$ }} &  \multirow{4}{*}{$q=2^4$} & \multirow{4}{*}{changed locality} \\
    & & & \\
    & & & \\
    & & & \\
    \hline
    \end{tabular}
    \label{compare construction}
\end{table}

\section{Preliminary}\label{sec pre}
We first introduce some notation used throughout this paper.
\begin{itemize}
\item For a positive integer $a$, let $[a]$ be the set $\{1,2,\dots,a\}$;
\item For a prime power $q$, let $\mathbb{F}_q$ be the finite field with $q$ elements and $\mathbb{F}_q^*=\mathbb{F}_q\setminus\{0\}$;
\item For two vector spaces $V_1,V_2$ over $\mathbb{F}_q$, denote $V_1 \otimes V_2 = \{(v_1,v_2) : v_1\in V_1, v_2\in V_2\}$;
\item For a finite set $S$, let $|S|$ denote its cardinality;
\item An $[n,k,d]_q$ linear code $\mathcal{C}$ is a $k$-dimensional subspace of $\mathbb{F}_q^n$ with the minimal Hamming distance $d$;
\item For a set $S\subseteq[n]$, let $\mathcal{C}|_S$ denote the punctured code formed by deleting the code symbols indexed by $[n]\setminus S$ from all codewords of $\mathcal{C}$;
\item Let $\bm{0}_{m\times n}$ and $I_{m\times n}$ denote the $m\times n$ zero matrix and identity matrix, respectively. If $m,n$ are clear from context, we simply write $\bm{0}$ and $I$ for short;
\item For a matrix
\begin{equation}\label{eqn20}
    H=\begin{pmatrix}
      A_1 & \bm{0} & \cdots & \bm{0} \\
      \bm{0} & A_2 & \cdots & \bm{0} \\
      \vdots & \vdots & \ddots & \vdots \\
      \bm{0} & \bm{0} & \cdots & A_g \\
      B_1 & B_2 & \cdots & B_g
    \end{pmatrix},
\end{equation}
let $H||_{\{i_1,i_2,\dots,i_t\}}$ denote the submatrix of $H$ as
\begin{equation}\label{eqn19}
   H||_{\{i_1,i_2,\dots,i_t\}}\triangleq \begin{pmatrix}
      A_{i_1} & \bm{0} & \cdots & \bm{0} \\
      \bm{0} & A_{i_2} & \cdots & \bm{0} \\
      \vdots & \vdots & \ddots & \vdots \\
      \bm{0} & \bm{0} & \cdots & A_{i_t} \\
      B_{i_1} & B_{i_2} & \cdots & B_{i_t}
    \end{pmatrix},
\end{equation}
where $1\le i_1<i_2<\dots<i_t\le g$;
\item Let $\top$ be the transpose operator.
\end{itemize}

It is well known that the following lemma characterizes the distance property of a punctured code.

\begin{lemma}[{\cite[Theorem 1.5.7]{huffman2010fundamentals}}]\label{lem4}
Let $\mathcal{C}$ be an $[n,k,d]_q$ code. For any set $S\subseteq[n]$, if $n-|S|<d$, then the punctured code $\mathcal{C}|_S$ has dimension $k$.
\end{lemma}

\subsection{Locally Repairable Codes}

\begin{definition}[\cite{gopalan2012locality},\cite{prakash2012optimal}]\label{def3}
Let $\mathcal{C}$ be an $[n,k,d]_q$ code. The $i$th code symbol of $\mathcal{C}$ is said to have $(r,\delta)$-locality if there exists a subset $S_i\subset[n]$ (an $(r,\delta)$-repair set) satisfying
\begin{itemize}
  \item $i\in S_i$ and $|S_i|\le r+\delta-1$,
  \item the minimum distance of the code $\mathcal{C}|_{S_i}$ is at least $\delta$.
\end{itemize}
\end{definition}

For an $[n,k,d]_q$ linear code $\mathcal{C}$, if all the $n$ code symbols have $(r,\delta)$-locality, then it is said to have all symbol $(r,\delta)$-locality, denoted as locally repairable codes (LRCs) with $(r,\delta)_a$-locality. For an LRC with $(r,\delta)_a$-locality, Lemma \ref{lem4} and Definition \ref{def3} imply that there exists a subset $\mathrm{Loc}(i)\subseteq S_i$ with $|\mathrm{Loc}(i)|\le r$ and $i\not\in\mathrm{Loc}(i)$, such that
\begin{equation*}
    c_i = \sum_{j\in \mathrm{Loc}(i)}\lambda_j c_j,
\end{equation*}
where $\lambda_j\in\mathbb{F}_q^*$ for all $j\in \mathrm{Loc}(i)$, for each codeword $\mathbf{c}=(c_1,c_2,\dots,c_n)\in\cC$. Note that, for given $i\in [n]$, $\mathrm{Loc}(i)$ may be not unique.  In the following, we use the notation $\mathrm{Loc}(i)$ to denote one of such sets for the $i$th code symbol.

Some upper bounds on the minimum Hamming distance of linear codes with $(r,\delta)_a$-locality were derived as follows.

\begin{lemma}[\cite{gopalan2012locality},\cite{prakash2012optimal}]\label{lem5}
The minimum distance of an $[n,k,d]_q$ code $\mathcal{C}$ with $(r,\delta)_a$-locality is upper bounded by
\begin{equation}\label{Sbound1}
d\le n-k+1-\left(\left\lceil\frac{k}{r}\right\rceil-1\right)(\delta-1).
\end{equation}
\end{lemma}

\begin{lemma}[{\cite[Remark II.4]{kong2021new}}]\label{lem1}
When $r=d-\delta$ and $(r+\delta-1)\mid n$, the Singleton-type bound \eqref{Sbound1} can be improved as
\begin{equation}\label{Sbound2}
d\le  n-k+1-\left\lceil\frac{k}{r}\right\rceil(\delta-1).
\end{equation}
\end{lemma}

Given $n,k,r,\delta,q$, let $d_{\opt}^{(q)}(n,k,r,\delta)$ is the largest possible minimum Hamming distance of an $[n,k]_q$ LRCs with $(r,\delta)_a$-locality for given finite field $\F_q$. When the bound in Lemma \ref{lem5} or Lemma \ref{lem1} is tight, then $d_{\opt}^{(q)}(n,k,r,\delta)=n-k+1-\left(\left\lceil\frac{k}{r}\right\rceil-1\right)(\delta-1)$ or $d_{\opt}^{(q)}(n,k,r,\delta)=n-k+1-\left\lceil\frac{k}{r}\right\rceil(\delta-1)$, respectively. An $[n,k,d]_q$ LRC with $(r,\delta)_a$-locality is called \textit{optimal} if $d=d_{\opt}^{(q)}(n,k,r,\delta)$. To the convenience of analysis, we assume that
\begin{equation}\label{eqn9}
d_{\opt}^{(q)}(n,k,r,\delta) = n-k+1-\phi(k,r,\delta),
\end{equation}
where $\phi(k,r,\delta)$ is determined by $k,r,\delta$.

The repair sets of linear code $\mathcal{C}$ with $(r,\delta)_a$-locality have the following important property.

\begin{lemma}[{\cite[Lemma 2]{cai2020optimal}}]\label{lem3}
Let $\mathcal{C}$ be an optimal linear LRC with $(r,\delta)_a$-locality and achieving the minimum Hamming distance bound in \eqref{Sbound1}. If $r\mid k$ and $r<k$, then there are $t$ repair sets $S_{i_1},\dots,S_{i_t}$ are mutually disjoint where $S_{i_j}\subseteq[n]$ for $j\in[t]$, and $|S_{i_j}|=r+\delta-1$ for all $j\in[t]$, and the punctured code $\mathcal{C}|_{S_{i_j}}$ is an $[r+\delta-1,r,\delta]$ MDS code. In particular, we have $(r+\delta-1)\mid n$, and $t=\frac{n}{r+\delta-1}$. Furthermore, the code $\mathcal{C}$ has a parity check matrix $H$ of the form \eqref{eqn20} where $g=t$ and $A_j$ is the parity check matrix of $\mathcal{C}|_{S_{i_j}}$ for $j\in[t]$ up to a rearrangement of the code coordinates.
\end{lemma}

\subsection{Convertible Codes}

Recently, \cite{ge2024mds} generalized the framework for studying code conversions proposed by \cite{maturana2022convertible}, allowing all initial and final codes to be two sets of codes with arbitrary parameters.

\begin{definition}[Irregular Conversion Procedure \cite{ge2024mds}]\label{def irregular cp}
Let $t_1,t_2$ be two positive integers. An irregular conversion procedure from $t_1$ initial codes $\mathcal{C}_{I_1},\mathcal{C}_{I_2},\cdots,\mathcal{C}_{I_{t_1}}$ to $t_2$ final codes $\mathcal{C}_{F_1},\mathcal{C}_{F_2},\cdots,\mathcal{C}_{F_{t_2}}$, where $\mathcal{C}_{I_i}$ is an $[n_{I_i},k_{I_i}]_q$ code with $r_{I_i}\triangleq n_{I_i}-k_{I_i}$ for $i\in[t_1]$,  and $\mathcal{C}_{F_j}$ is an $[n_{F_j},k_{F_j}]_q$ code with $r_{F_j}\triangleq  n_{F_j}-k_{F_j}$ for $j\in[t_2]$ satisfying
\begin{equation}\label{equ:dimension}
    \sum_{i\in[t_1]}k_{I_i}=\sum_{j\in[t_2]}k_{F_j},
\end{equation}
is a function $\sigma = (\sigma_1,\sigma_2,\cdots,\sigma_{t_2})$  from $\bigotimes_{i\in [t_1]}\mathcal{C}_{I_{i}} \rightarrow \bigotimes_{j\in [t_2]}\mathcal{C}_{F_j}$, i.e.,
$$ \sigma_j : \bigotimes_{i\in[t_1]} \mathcal{C}_{I_i} \rightarrow \mathcal{C}_{F_j},\quad 1\leq j\leq t_2. $$

For the convenience of analysis, we use two-dimensional indices to
denote the code symbols, i.e., the codeword $\bm{c}_{I_i}=(c_{i,1},c_{i,2},$ $\cdots, c_{i,n_{I_i}})\in \mathcal{C}_{I_i}$ for $i\in [t_1]$. A concrete conversion procedure is realized in three steps as follows:

\textbf{Step 1: Choosing unchanged symbols from initial codes.}  For $j\in [t_2]$, the conversion procedure keeps some code symbols from each initial code $\mathcal{C}_{I_i},1\le i\le t_1$. Denote the indices of these symbols as  $\unchanged_{1,j},\unchanged_{2,j},\cdots,\unchanged_{t_1,j}$, where $\unchanged_{i,j}\subseteq\{i\}\times[n_{I_i}]$ satisfies
 \begin{equation*}
 \unchanged_{i,\tau_1}\cap \unchanged_{i,\tau_2}=\emptyset \,\, \text{ for }\tau_1\ne \tau_2\in[t_2].
\end{equation*}

\textbf{Step 2: Reading symbols from  initial codes.} For $j\in [t_2]$, the function $\sigma_j$ reads some code symbols from each initial code $\mathcal{C}_{I_i},1\le i\le t_1$, whose indices are denoted by $\reading_{1,j},\reading_{2,j},\cdots,\reading_{t_1,j}$ where $\reading_{i,j}\subseteq\{i\}\times[n_{I_{i}}]$, respectively.

\textbf{Step 3: Conversion.} For any $j\in [t_2]$, choose $u_j$ unchanged symbols $\mathcal{C}_{I_1}|_{\unchanged_{1,j}},\mathcal{C}_{I_2}|_{\unchanged_{2,j}},\cdots,\mathcal{C}_{I_{t_1}}|_{\unchanged_{t_1,j}}$. And then  apply  $\sigma_j$ to the read symbols $\mathcal{C}_{I_1}|_{\reading_{1,j}},\mathcal{C}_{I_2}|_{\reading_{2,j}},\cdots,\mathcal{C}_{I_{t_1}}|_{\reading_{t_1,j}}$ to generate
$n_{F_j}-u_j$ symbols
\begin{equation}\label{Eqn_Gen_Fj}
\mathcal{C}_{F_j}|_{\written_{j}}
= \left\{ \sigma_j(\bm{x}) : \bm{x}\in\bigotimes_{i\in[t_1]}\mathcal{C}_{I_i}|_{\reading_{i,j}} \right\},
\end{equation}
 which are called written symbols, where $\written_{j}\triangleq \{w_j\}\times[n_{F_j}-u_j]$ and $w_j\triangleq t_1+j$.

Finally, the final code $\mathcal{C}_{F_j}$ is given by
\begin{align}
\mathcal{C}_{F_j}
&=\left\{ (\bm{x}_1,\sigma_j(\bm{x}_2)) : \bm{x}\in\bigotimes_{i\in[t_1]}\mathcal{C}_{I_i}, \bm{x}_1=\bm{x}|_{\bigcup_{i\in[t_1]}\unchanged_{i,j}}, \bm{x}_2=\bm{x}|_{\bigcup_{i\in[t_1]}\reading_{i,j}} \right\} \label{eqn_5}\\
&\subseteq \bigotimes_{i\in[t_1]}\mathcal{C}_{I_i}|_{\unchanged_{i,j}} \otimes \mathcal{C}_{F_j}|_{\written_j} \notag
\end{align}
for each $1\le j\le t_2$, where we assume the code symbols of $\mathcal{C}_{F_j}$ are
indexed by the set
\begin{equation}\label{eqn18}
\bigcup_{i\in[t_1]}\unchanged_{i,j}\cup\written_{j}
\end{equation}
such that
\begin{equation}\label{eqn_1}
    \mathcal{C}_{I_i}|_{\unchanged_{i,j}} = \mathcal{C}_{F_j}|_{\unchanged_{i,j}}.
\end{equation}
\end{definition}

\begin{definition}[Irregular Convertible Code \cite{ge2024mds}]\label{def_2}
Let $t_1,t_2$ be two positive integers. A $(t_1,t_2)_q$ irregular convertible code $\bm{\mathcal{C}}$ over $\mathbb{F}_q$  consists of: (1) $t_1$ initial codes $\mathcal{C}_{I_1},\mathcal{C}_{I_2},\cdots,\mathcal{C}_{I_{t_1}}$ and $t_2$ final codes $\mathcal{C}_{F_1},\mathcal{C}_{F_2},\cdots,\mathcal{C}_{F_{t_2}}$, where $\mathcal{C}_{I_i}$ is an $[n_{I_i},k_{I_i}]_q$ code for $i\in[t_1]$,  and $\mathcal{C}_{F_j}$ is an $[n_{F_j},k_{F_j}]_q$ code for $j\in[t_2]$ satisfying $\sum_{i\in[t_1]}k_{I_i} = \sum_{j\in[t_2]}k_{F_j}$; (2) an irregular conversion procedure $\sigma$ defined by Definition \ref{def irregular cp}.
\end{definition}

When both the initial and final codes are MDS codes, the corresponding convertible code $\bm{\mathcal{C}}$ is called an \textit{MDS convertible code}. When both the initial and final codes are LRCs, the corresponding convertible code $\bm{\mathcal{C}}$ is called an \textit{locally repairable convertible code} (LRCC). In this paper, we mainly consider the case where all initial codes have the same parameters in the merge regime, i.e., code length $n_I$, dimension $k_I$ and $(r_I,\delta_I)$-locality, which is called a $(t,1)_q$ LRCC with parameters $(n_I,k_I,r_I,\delta_I;n_F,k_F,r_F,\delta_F)_a$, where the subscript $a$ means both the initial and final codes are with all symbols locality. If some parameters $r_I,\delta_I$ are unknown, then we say the above LRCC is with parameters $(n_I,k_I,*,*;n_F,k_F,r_F,\delta_F)_a$.

One merit factor of a convertible code is the read and write access costs, which measure the number of symbols accessed by the conversion procedure.

\begin{definition}[Access Cost \cite{maturana2022convertible}\cite{ge2024mds}]
For a $(t_1,t_2)_q$ irregular convertible code $\bm{\mathcal{C}}$, the access cost  $\rho$ is defined as the sum of read access cost $\rho_r=\sum_{i\in[t_1]}|\bigcup_{j\in[t_2]}\reading_{i,j}|$ and write access cost $\rho_w=\sum_{j\in[t_2]}(n_{F_j}-\sum_{i\in[t_1]}|\mathcal{U}_{i,j}|)$, where the read access cost is the number of symbols in initial codes used in the conversion procedure functions and the write access cost counts the number of symbols written in the final codes.
\end{definition}

In \cite{kong2024locally}, Kong proposed a lower bound on the access cost of an LRCC with the condition that the final code $\mathcal{C}_F$ is an $[n,k,d]_q$-LRC with $(r,\delta)_a$-locality.

\begin{theorem}[{\cite[Theorem IV.4]{kong2024locally}}]\label{thm1}
Let $\bm{\mathcal{C}}$ be a $(t,1)_q$ convertible code with parameters $(n_I,k,*,*;n_F,tk,r,\delta)_a$. Assume that $\mathcal{C}_F$ is a linear code with minimum distance $d$. Then, the write access cost is
\begin{equation*}
    \rho_w \ge t\left( d+(t-1)k+ \left\lceil \frac{(t-1)k}{r} \right\rceil -2 \right) - (t-1)n_F.
\end{equation*}
Write $\overline{\Delta}=n_F-2d-((t-1)k+\lceil\frac{(t-1)k}{r}\rceil)+3$. Then, the read access cost is
\begin{equation*}
    \rho_r \ge
    \begin{cases}
      tk, & \mbox{if } \overline{\Delta}\le0 \text{ or } d>n_I-k+1, \\
      t(k-\lceil\frac{r\overline{\Delta}}{r+1}\rceil), & \mbox{otherwise}.
    \end{cases}
\end{equation*}
\end{theorem}

\begin{remark}
  When $r=k$, the preceding bound is equivalent to the lower bound on the access cost for MDS convertible codes proposed by \cite[Theorem 9]{maturana2022convertible}, since the $[n,k,d]_q$ MDS codes are optimal LRCs with $(k,d)_a$-locality.
\end{remark}

\section{Improved Lower Bound on Access Cost}\label{sec bound}
In this section, we improve Theorem \ref{thm1} by establishing a new lower bound that
depends on both $r$ and $\delta$.
At first, we derive an upper bound on the number of unchanged symbols for each initial code.

\begin{lemma}\label{lem6}
Let $\bm{\mathcal{C}}$ be a $(t,1)_q$ LRCC with parameters $(n_I,k,*,*;n_F,tk,r,\delta)_a$. Assume that the final code $\mathcal{C}_F$ has the minimum distance $d$ and the optimal Hamming distance $d_{\opt}^{(q)}(n,k,r,\delta) = n-k+1-\phi(k,r,\delta)$. Then, for every $1\le i\le t$,
\begin{equation*}
|\unchanged_{i,1}| \le n_F - d - (t-1)k + 1 -\phi((t-1)k,r,\delta).
\end{equation*}
\end{lemma}
\begin{IEEEproof}
Without loss of generality, assume $i=1$. Let \( \tilde{\mathcal{C}} \) be the subcode of the final code \( \mathcal{C}_F \) that is obtained by converting the initial codewords \( \bm{c}_1, \bm{c}_2, \dots, \bm{c}_t \), where \( \bm{c}_1 = \bm{\alpha}\in \mathcal{C}_{I_1} \) is a constant vector and \( \bm{c}_i \in \mathcal{C}_{I_i} \) for $i\in\{2,\dots,t\}$. Let $d(\tilde{\mathcal{C}})$ denote the minimum Hamming distance of $\tilde{\mathcal{C}}$. Then we have $d(\tilde{\mathcal{C}})\ge d$ and $|\tilde{\mathcal{C}}|=q^{(t-1)k}$. Define
\begin{equation*}
    N \triangleq \{ i\in\mathcal{F}_1 : i\in \unchanged_{1,1} \text{ or exist a set } \mathrm{Loc}(i)\subseteq\unchanged_{1,1} \},
\end{equation*}
where $\mathcal{F}_1$ is defined by \eqref{eqn18}.

Let $\bm{v}=\sigma_1(\bm{\alpha},\bm{c}_2,\dots,\bm{c}_t) \in\tilde{\mathcal{C}}$ be the codeword converted from codewords $\bm{\alpha},\bm{c}_2,\dots,\bm{c}_t$, where $\sigma_1$ is the conversion procedure
defined in Definition \ref{def_2}. Note that $\bm{v}|_{\mathcal{U}_{1,1}} = \bm{\alpha}|_{\mathcal{U}_{1,1}}$, and symbols in $\bm{v}|_N$ can be recovered by symbols in $\bm{v}|_{\mathcal{U}_{1,1}}$. Thus, $\bm{v}|_N$ is fully determined by $\bm{\alpha}|_{\mathcal{U}_{1,1}}$. This implies that $\tilde{\mathcal{C}}|_N$ is a constant vector determined by $\bm{\alpha}|_{\mathcal{U}_{1,1}}$. Therefore, we have $d(\tilde{\mathcal{C}})=d(\tilde{\mathcal{C}}|_{N^C})$ and $|\tilde{\mathcal{C}}| = |\tilde{\mathcal{C}}|_{N^C}| = q^{(t-1)k}$.

On the other hand, for each $i\in N^C \triangleq \mathcal{F}_1\setminus N$ and its corresponding $\mathrm{Loc}(i)$, we claim that $\mathrm{Loc}(i)\cap N^C \ne \emptyset$. Otherwise, assume that there is a $j\in N^C$ such that $\mathrm{Loc}(j)\subseteq N$. Then, each symbol whose index is in $\mathrm{Loc}(j)$ can be recovered by symbols with indices in $\mathcal{U}_{1,1}$, implying that the $j$th symbol can be recovered by symbols with indices in $\mathcal{U}_{1,1}$. This leads to $j\in N$, a contradiction.

Since for each \( i \in N^C \), \( \mathrm{Loc}(i) \) satisfies \( \mathrm{Loc}(i) \cap N^C \neq \emptyset \), and \( \mathrm{Loc}(i) \subseteq S_i \), where \( S_i \) is an \( (r,\delta) \)-repair set for the code symbol \( c_i \), we have
\( d(\tilde{\mathcal{C}}|_{S_i}) \ge d(\mathcal{C}_F|_{S_i})\ge\delta \) and \( \tilde{\mathcal{C}}|_{S_i\cap N} \) is a constant vector. It follows that $$ d(\tilde{\mathcal{C}}|_{N^C\cap S_i}) = d(\tilde{\mathcal{C}}|_{S_i}) \ge \delta.$$
 Hence, \( \tilde{\mathcal{C}}|_{N^C} \) can be regarded as an \( (n_F-|N|,(t-1)k) \)-LRC with \( (r,\delta)_a \)-locality. We obtain
\begin{equation}\label{eqn15}
\begin{split}
   d(\tilde{\mathcal{C}}) &= d(\tilde{\mathcal{C}}|_{N^C}) \\
     &\le d_{\opt}^{(q)}(n_F-|N|,(t-1)k,r,\delta) \\
     &\le (n_F-|N|)-(t-1)k+1-\phi((t-1)k,r,\delta) .
\end{split}
\end{equation}
The desirable result holds from $d(\tilde{\mathcal{C}})\ge d$, $|\mathcal{U}_{1,1}|\le |N|$ and \eqref{eqn15}.
\end{IEEEproof}

Secondly, we provide a lower bound on the number of read symbols from each initial code.

\begin{lemma}\label{lem7}
Let $\bm{\mathcal{C}}$ be a $(t,1)_q$ LRCC with parameters $(n_I,k,*,*;n_F,tk,r,\delta)_a$. For $1\le i\le t$, define
\begin{equation}\label{eqn_Delta_i}
\Delta_i\triangleq |\mathcal{U}_{i,1}\setminus \mathcal{D}_{i,1}|-d+1.
 \end{equation}
 Let $B=\cup_{1\leq t\leq \tau} B_t$ be the subset of $\mathcal{U}_{i,1}\setminus \mathcal{D}_{i,1}$ with size at most $  (\delta-1) \lfloor \frac{\Delta_i}{r+\delta-1} \rfloor$ for a positive integer $\tau$, such that
 \begin{itemize}
   \item[(i)] {For each $1\leq t\leq \tau$,  all code symbols indexed by $i\in B_t$ are in the same repair set $S_t$ of the final code $\cC_F$, and $B_{t_1} \cap S_{t_2} = \emptyset$ for all $1\leq t_2 < t_1 \leq \tau$;
   }
   \item[(ii)] {For each $1\leq t\leq \tau$, $|B_t|\le \delta-1$, and
       \begin{equation}\label{eqn_repair}
       \left| \left( S_{t} \setminus \bigcup_{1\leq j\leq t-1} S_j \right) \cap (\cU_{i,1} \setminus \reading_{i,1}) \right| < \delta-1
        \mbox{ when } |B_t| < \delta-1.
       \end{equation}
       }
 \end{itemize}
  Then,
\begin{equation*}
 |\mathcal{D}_{i,1}| \ge \begin{cases}
             k, & \mbox{if } \Delta_i\le0, \\
             k-\Delta_i+|B|, & \mbox{otherwise}.
           \end{cases}
\end{equation*}
Moreover, if $d>n_I-k+1$, then $\Delta_i\le0$ for $1\le i\le t$ and therefore, $|\mathcal{D}_{i,1}|\ge k$.
\end{lemma}
\begin{IEEEproof}
Without loss of generality, assume \( i = 1 \). Let \( \tilde{\mathcal{C}} \) be the subcode of the final code \( \mathcal{C}_F \) that arises from converting the initial codewords \( \bm{c}_1, \bm{c}_2, \dots, \bm{c}_t \), where \( \bm{c}_i = \bm{\beta}_i \) are constant vectors for \( i = 2, \dots, t \). Let \( d(\tilde{\mathcal{C}}) \) denote the minimum distance of \( \tilde{\mathcal{C}} \). Then we have \( d(\tilde{\mathcal{C}}) \geq d \) and \( |\tilde{\mathcal{C}}| = q^k \).

For any \( \bm{v} \in \tilde{\mathcal{C}} \), each symbol of \( \bm{v} \) is either inherited from \( \bm{c}_1|_{\mathcal{U}_{1,1}}, \bm{\beta}_2|_{\mathcal{U}_{2,1}}, \dots, \bm{\beta}_t|_{\mathcal{U}_{t,1}} \) or determined by symbols from \( \bm{c}_1|_{\mathcal{D}_{1,1}}, \bm{\beta}_2|_{\mathcal{D}_{2,1}}, \dots, \bm{\beta}_t|_{\mathcal{D}_{t,1}} \). Therefore, the non-constant symbols of the codewords in \( \tilde{\mathcal{C}} \) are either inherited from \( \bm{c}_1|_{\mathcal{U}_{1,1} \setminus \mathcal{D}_{1,1}} \) or are functions of symbols in \( \bm{c}_1|_{\mathcal{D}_{1,1}} \).

According to Lemma \ref{lem4}, we will pick a subset $M\subseteq\mathcal{F}_1$ of size at least $n_F-d+1$ and prove the lower bound by analyzing the relationships among codeword symbols in $\bm{c}_1|_{\mathcal{U}_{1,1} \setminus \mathcal{D}_{1,1}}$, $\bm{c}_1|_{\mathcal{D}_{1,1}}$ and $\bm{v}|_M$.

\textbf{Case 1}: $\Delta_1\le0$, i.e., $|\mathcal{U}_{1,1} \setminus \mathcal{D}_{1,1}| \le d-1$ by \eqref{eqn_Delta_i}.
In this case, $\mathcal{F}_1\setminus (\mathcal{U}_{1,1}\setminus \mathcal{D}_{1,1})$ has size at least $n_F-d+1\ge n_F-d(\tilde{\mathcal{C}})+1$ by $|\mathcal{F}_1|=n_F$. Take $M=\mathcal{F}_1\setminus (\mathcal{U}_{1,1} \setminus \mathcal{D}_{1,1})$, we have $|\tilde{\mathcal{C}}|_M|=|\tilde{\mathcal{C}}|=q^k$. Note that $\bm{c}_1|_{\mathcal{U}_{1,1}\setminus \mathcal{D}_{1,1}} = \bm{v}|_{\mathcal{U}_{1,1} \setminus \mathcal{D}_{1,1}}$. Thus, by $M\cap (\mathcal{U}_{1,1} \setminus \mathcal{D}_{1,1})=\emptyset$, non-constant symbols in $\bm{v}|_M$ are functions of symbols in $\bm{c}_1|_{\mathcal{D}_{1,1}}$. Therefore, $|\tilde{\mathcal{C}}|_M|$ is upper bounded by $q^{|\mathcal{D}_{1,1}|}$, which implies that $|\mathcal{D}_{1,1}|\ge k$.

\textbf{Case 2}: $\Delta_1>0$, i.e., $|\mathcal{U}_{1,1} \setminus \mathcal{D}_{1,1}| > d-1$ by \eqref{eqn_Delta_i}. By condition (ii), $|B_i| \le \delta-1$. Denote $\cA_1 \triangleq \{ i: |B_i|=\delta-1, i\in[\tau] \}$ and $\cA_2 \triangleq \{ i: |B_i|<\delta-1, i\in[\tau] \}$. Thus, $B = \cup_{i\in \cA_1 \cup \cA_2} B_i$. Let $$N = \bigcup_{ i\in \cA_1} (S_i \setminus B_i).$$
Thus,
\begin{equation*}
|B\cup N|
\leq |B| + |\cA_1|r
\leq \left\lfloor\frac{\Delta_1}{r+\delta-1} \right\rfloor \cdot (r+\delta-1)
\leq \Delta_1,
\end{equation*}
where the first inequality holds by $|S_t| \le r+\delta-1$ for $t\in[\tau]$, and the second inequality comes from condition (i) that $B_{t_1} \cap S_{t_2} = \emptyset$ for $1\leq t_2 < t_1 \leq \tau$, which implies $|\cA_1| \le \lfloor \frac{\Delta_1}{r+\delta-1} \rfloor$ by $|B| \le  (\delta-1) \lfloor \frac{\Delta_1}{r+\delta-1} \rfloor$. Note that
\begin{equation*}
    |(\mathcal{U}_{1,1}\setminus \mathcal{D}_{1,1})\setminus (B\cup N)|
    \ge |\mathcal{U}_{1,1}\setminus \mathcal{D}_{1,1}| - |B\cup N|
    \ge |\mathcal{U}_{1,1}\setminus \mathcal{D}_{1,1}| - \Delta_1
    = d-1.
\end{equation*}
Therefore, we can obtain an $(n_F-d+1)$-subset $M\subseteq\mathcal{F}_1$ by removing any $d-1$ elements in $(\mathcal{U}_{1,1}\setminus \mathcal{D}_{1,1}) \setminus (B\cup N)$ from $\mathcal{F}_1$. Note that both $B\cup N$ and $\mathcal{F}_1 \setminus (\mathcal{U}_{1,1}\setminus \mathcal{D}_{1,1})$ are subsets of $M$. Thus, by $n_F-d+1\ge n_F-d(\tilde{\mathcal{C}})+1$, we have $|\tilde{\mathcal{C}}|_M|=q^k$.

\textbf{Claim 1}: We claim that the symbols $\bm{v}|_{B_1\cup \cdots \cup B_{\ell}}$ can be recovered by the symbols $\bm{v}|_{M\setminus (B_1\cup \cdots \cup B_{\ell})}$ for $1\leq \ell \leq \tau$. The claim can be proved by induction as follows. Denote $\overline{\cU} \triangleq \cU_{2,1} \cup \cdots \cup \cU_{t,1}$.
If $\ell=1\in \cA_2$, then $|S_1 \cap (\cU_{1,1}\setminus \reading_{1,1})|<\delta-1$ by \eqref{eqn_repair}.
By \eqref{eqn18} and the fact $S_1$ is a repair set in $\mathcal{C}_{F}$, we have $S_1 \subseteq \cF_1 = \cU_{1,1} \cup \overline{\cU} \cup \cW_1$. Furthermore, we have
\begin{align*}
& S_1 \cap ((\cU_{1,1} \cap \reading_{1,1}) \cup \overline{\cU} \cup \cW_1) \\
=~& S_1 \cap (\cF_1 \setminus (\cU_{1,1} \setminus \reading_{1,1})) \\
\end{align*}
and
\begin{align*}
& |S_1 \cap ((\cU_{1,1} \cap \reading_{1,1}) \cup \overline{\cU} \cup \cW_1)| \\
=~& |S_1 \cap (\cF_1 \setminus (\cU_{1,1} \setminus \reading_{1,1}))| \\
=~&|S_1|-|S_1 \cap (\cU_{1,1}\setminus \reading_{1,1})|\\
>~& |S_1|-\delta+1.
\end{align*}
 Thus, by $(r,\delta)$-locality and the fact $S_1$ is a repair set, we obtain that
 $\bm{v}|_{B_1}$ can be recovered by $\bm{v}|_{\cF_1 \setminus (\cU_{1,1} \setminus \reading_{1,1})}$ and
 then can be recovered by $\bm{v}|_{M\setminus B_1}$ since
 $$\cF_1 \setminus (\cU_{1,1} \setminus \reading_{1,1})\subseteq M,$$
 $$B_1\subseteq M,\quad B_1\subseteq \cU_{1,1} \setminus \reading_{1,1},$$
 i.e.,
 $$B_1\cap (\cF_1 \setminus (\cU_{1,1} \setminus \reading_{1,1}))=\emptyset.$$
If $\ell=1\in\cA_1$, the repair set $S_1 \subseteq B \cup N \subseteq M$, therefore, the symbols $\bm{v}|_{B_1}$ can be recovered by the symbols $\bm{v}|_{M\setminus B_1}$ by condition (ii) that $|B_1|\le \delta-1$.
In both subcases, we have $v|_{S_1}$ can also be recovered by $\bm{v}|_{M\setminus B_1}$.

Assume that the symbols $\bm{v}|_{B_1\cup \dots\cup B_\ell}$ and $\bm{v}|_{S_1\cup \dots\cup S_\ell}$ can be recovered by the symbols $\bm{v}|_{M\setminus (B_1\cup \dots\cup B_\ell)}$ for $1\leq \ell < \tau$. Consider the case $\ell+1$. Since condition (i) that $B_{t_1} \cap S_{t_2} = \emptyset$ for all $1 \leq t_2 < t_1 \leq \tau$, then  the symbols $\bm{v}|_{B_1\cup \dots\cup B_\ell}$ and $\bm{v}|_{S_1\cup \dots\cup S_\ell}$  can be recovered by the symbols $\bm{v}|_{M\setminus (B_1\cup \dots\cup B_{\ell+1})}$ by our assumption. It is sufficient to show that the symbols $\bm{v}|_{B_{\ell+1}}$ and $\bm{v}|_{S_{\ell+1}}$ can be recovered by the symbols $\bm{v}|_{M\setminus B_{\ell+1}}$ since the symbols $\bm{v}|_{B_1\cup \dots\cup B_\ell}$ are recoverable in this case.

For the subcase $\ell+1 \in \cA_2$, by \eqref{eqn_repair}, we have
      \begin{equation}\label{eqn29}
        \left| \left( S_{\ell+1} \setminus \bigcup_{i\in[\ell]} S_i \right) \cap (\cU_{1,1} \setminus \reading_{1,1}) \right| < \delta-1,
      \end{equation}
which implies
\begin{align*}
&  \left| \left( S_{\ell+1} \cap \left(M \setminus B_{\ell+1} \right) \right) \cup \left( \bigcup_{i\in[\ell]} (S_i \cap S_{\ell+1}) \right) \right| \\
\ge~& \left| ( S_{\ell+1} \setminus (\cU_{1,1} \setminus \reading_{1,1}))  \cap (M \setminus B_{\ell+1} ) ) \cup \left( \bigcup_{i\in[\ell]} (S_i \cap S_{\ell+1}) \right) \right| \\
=~&  \left| ( S_{\ell+1} \setminus (\cU_{1,1} \setminus \reading_{1,1}) ) \cup \left( \bigcup_{i\in[\ell]} (S_i \cap S_{\ell+1}) \right) \right| \\
=~& \left| S_{\ell+1} \setminus \left( \left( S_{\ell+1} \setminus \bigcup_{i\in[\ell]} S_i \right) \cap (\cU_{1,1} \setminus \reading_{1,1}) \right) \right| \\
\geq~& |S_{\ell+1}|-\delta+1,
\end{align*}
where the first equation holds by $S_{\ell+1} \setminus (\cU_{1,1} \setminus \reading_{1,1}) \subseteq M$ and $B_{\ell+1} \subseteq \cU_{1,1} \setminus \reading_{1,1}$, i.e., $( S_{\ell+1} \setminus (\cU_{1,1} \setminus \reading_{1,1}))  \subseteq (M \setminus B_{\ell+1} )$, the last inequality comes from \eqref{eqn29}. Then, the symbols $\bm{v}|_{B_{\ell+1}}$ can be recovered by the symbols $\bm{v}|_{M \setminus B_{\ell+1}}$ according to the $(r,\delta)$-locality.

If $\ell+1 \in \cA_1$, then $S_{\ell+1} \subseteq B \cup N \subseteq M$. Since the symbols $\bm{v}|_{S_1\cup \dots\cup S_\ell}$ are known, we have that
      \begin{align*}
        &\left| (S_{\ell+1} \cap (M \setminus B_{\ell+1})) \cup \left( \bigcup_{i\in[\ell]} (S_i \cap S_{\ell+1}) \right) \right| \\
        =~& \left|(S_{\ell+1} \setminus B_{\ell+1}) \cup \left( \bigcup_{i\in[\ell]} (S_i \cap S_{\ell+1}) \right) \right| \\
        \ge~& \left|S_{\ell+1} \setminus B_{\ell+1} \right| \\
        \ge~& |S_{\ell+1}| -\delta+1,
      \end{align*}
      where the first equation holds by $B_{\ell+1} \subseteq S_{\ell+1} \subseteq B \cup N \subseteq M$, the first inequality comes from the condition (i) that $B_{\ell+1} \cap S_i = \emptyset$ for $i\in[\ell]$, and the second inequality holds by condition (ii) that $|B_{\ell+1}| \le \delta-1$.  Then, the symbols $\bm{v}|_{B_{\ell+1}}$ and $\bm{v}|_{S_{\ell+1}}$ can be recovered by the symbols $\bm{v}|_{M \setminus B_{\ell+1}}$.
Therefore, the symbols $\bm{v}|_{B_1\cup \cdots \cup B_{\tau}} = \bm{v}|_{B}$ can be recovered by $\bm{v}|_{M\setminus (B_1\cup \cdots \cup B_{\tau})} = \bm{v}|_{M\setminus B}$ according to induction.

Let $M_1=M\cap (\mathcal{U}_{1,1}\setminus \mathcal{D}_{1,1})$. Then, $|M_1|=|(\mathcal{U}_{1,1}\setminus \mathcal{D}_{1,1})|-d+1=\Delta_1$.
Recall that $B\subseteq \mathcal{U}_{1,1}\setminus \mathcal{D}_{1,1}$ and $B\cup N\subseteq M$.
Since $\bm{v}|_{M\setminus(\cU_{1,1}\setminus \reading_{1,1})}$ are determined by $\bm{c}_1|_{\reading_{1,1}}$, thus the non-constant symbols in $\bm{v}|_{M\setminus(\cU_{1,1}\setminus \reading_{1,1})}$ and $\bm{v}|_{M_1\setminus B}$ (i.e., $ \bm{v}|_{M\setminus B}$) are determined by \( \bm{c}_1|_{\mathcal{D}_{1,1}} \) and \( \bm{c}_1|_{M_1\setminus B} \).
Now, by  the above Claim 1, we have  the non-constant symbols in \( \bm{v}|_{M} \) are determined by \( \bm{c}_1|_{\mathcal{D}_{1,1}} \) and \( \bm{c}_1|_{M_1\setminus B} \).  Therefore, \( |\tilde{\mathcal{C}}|_M| \) is upper bounded by
\[
q^{|\mathcal{D}_{1,1} \cup (M_1 \setminus B)|} = q^{|\mathcal{D}_{1,1}| + |M_1| - |B|},
\]
where the equality follows from \( M_1 \setminus B \subseteq \mathcal{U}_{1,1} \setminus \mathcal{D}_{1,1} \), \( (\mathcal{U}_{1,1} \setminus \mathcal{D}_{1,1}) \cap \mathcal{D}_{1,1} = \emptyset \), and \( B \subseteq M_1 \). This leads to \( |\mathcal{D}_{1,1}| + |M_1| - |B| \geq k \), which further implies that \( |\mathcal{D}_{1,1}| \geq k - \Delta_1 + |B| \).

\textbf{Case 3}: $d>n_I-k+1$. We claim that one must have $\Delta_1\le 0$. Otherwise, assume that $\Delta_1>0$ and let $Q\subseteq \mathcal{U}_{1,1}\setminus \mathcal{D}_{1,1}$ such that $|Q|=d-1$. Let $M=\mathcal{F}_1\setminus Q$. By $|M|=n_F-d+1\ge n_F-d(\tilde{\mathcal{C}})+1$, we have $|\tilde{\mathcal{C}}|_M|=q^k$. Note that for each $\bm{v}\in\tilde{\mathcal{C}}$, the non-constant symbols of $\bm{v}|_M$ are either inherited from $\bm{c}_1|_{\mathcal{U}_{1,1}\setminus Q}$ or determined by $\bm{c}_1|_{\mathcal{D}_{1,1}}$. Therefore, we have $|\tilde{\mathcal{C}}|_M|=q^k\le q^{|(\mathcal{U}_{1,1}\setminus Q)\cup \mathcal{D}_{1,1}|}$. Since $Q\subseteq \mathcal{U}_{1,1}\setminus \mathcal{D}_{1,1}$, we have $(\mathcal{U}_{1,1}\setminus Q)\cup \mathcal{D}_{1,1}=(\mathcal{U}_{1,1}\cup \mathcal{D}_{1,1})\setminus Q$ and $|(\mathcal{U}_{1,1}\cup \mathcal{D}_{1,1})\setminus Q| = |\mathcal{U}_{1,1}\cup \mathcal{D}_{1,1}|-|Q|$. This leads to
\begin{equation*}
    |(\mathcal{U}_{1,1}\setminus Q)\cup \mathcal{D}_{1,1}| = |\mathcal{U}_{1,1} \cup \mathcal{D}_{1,1}|-|Q| \ge k.
\end{equation*}
On the other hand, since $\mathcal{U}_{1,1}\cup \mathcal{D}_{1,1}\subseteq \{1\}\times[n_{I}]$, we have
\begin{equation*}
    k\leq |\mathcal{U}_{1,1}\cup \mathcal{D}_{1,1}|-|Q|\le n_I-d+1.
\end{equation*}
This implies that $d\le n_I-k+1$, which is a contradiction. Thus, when $d>n_I-k+1$, we always have $\Delta_1\le 0$ and thus $|\mathcal{D}_{1,1}|\ge k$, which completes the proof.
\end{IEEEproof}

%\begin{remark}
%For $\delta=2$, condition (ii) of Lemma \ref{lem7} can only be satisfied  when $|B_i|=1$ for all $1\leq i\leq \tau$.
%\end{remark}

The lower bound in Lemma \ref{lem7} depends critically on the size of the set $B$. Below, we demonstrate that there always exists a set $B$ with large cardinality.

\begin{lemma}\label{lem8}
Assume that $\mathcal{C}$ is a linear $[n_F,tk]_q$ LRC with $(r,\delta)_a$-locality. For any $\tilde{V}\subseteq[n_F]$, there is a subset $B=\cup_{1\leq t\leq \tau} B_t \subseteq \tilde{V}$ of size $(\delta-1)\lfloor\frac{|\tilde{V}|}{r+\delta-1}\rfloor$ for a positive integer $\tau$, such that
 \begin{itemize}
   \item[(i)] {For each $1\leq t\leq \tau$, all code symbols indexed by $i\in B_t$ are in the same repair set $S_t$ of the code $\cC$, and $B_{t_1} \cap S_{t_2} = \emptyset$ for all $1\leq t_2 < t_1 \leq \tau$;
   }
   \item[(ii)] {For each $1\leq t\leq \tau$, $|B_t|\le \delta-1$, and
       \begin{equation}\label{eqn30}
         \left| \left( S_{t} \setminus \bigcup_{1\leq i\leq t-1} S_i \right) \cap \tilde{V} \right| < \delta-1 \mbox{ when } |B_t| < \delta-1.
       \end{equation}
       }
 \end{itemize}
\end{lemma}
\begin{IEEEproof}
We find a subset $V$ of $\tilde{V}$ by the following process.
For $\ell=1$ and $i_1\in V^{(1)} = \tilde{V}$, let $\tilde{V}_{i_1}=V^{(1)} \cap S_{i_1}$,
\begin{equation}\label{eqn_B_1}
V_{i_1}
=\begin{cases}
V^*, \text{ where } V^*\subseteq \tilde{V}_{i_1} \text{ and } |V^*|=\delta-1, & \text{ if }|\tilde{V}_{i_1}|\geq \delta-1,\\
\tilde{V}_{i_1}, &\text{ if }|\tilde{V}_{i_1}|< \delta-1,
\end{cases}
\end{equation}
and
set
\begin{equation}\label{eqn_new_B1}
V^{(2)} = V^{(1)} \setminus S_{i_1}.
\end{equation}
For $1<\ell \leq |\tilde{V}|$, if $V^{(\ell)} \ne \emptyset$
then we set $\tilde{V}_{i_{\ell}}=V^{(\ell)} \cap S_{i_\ell}$  for an $i_\ell\in V^{(\ell)}$, and let
\begin{equation}\label{eqn_B_tau}
V_{i_\ell}
=\begin{cases}
V^*, \text{ where } V^*\subseteq \tilde{V}_{i_\ell} \text{ and } |V^*|=\delta-1, & \text{ if }|\tilde{V}_{i_\ell}|\geq \delta-1,\\
\tilde{V}_{i_{\ell}}, &\text{ if }|\tilde{V}_{i_{\ell}}|< \delta-1,
\end{cases}
\end{equation}
and then set
\begin{equation}\label{eqn_new_B2}
V^{(\ell+1)} = V^{(\ell)} \setminus S_{i_\ell}.
\end{equation}
For $1<\ell \leq |\tilde{V}|$, if $V^{(\ell)}=\emptyset$, then we get $\ell-1$ pairwise disjoint subsets  $V_{i_1}, V_{i_2}, \dots, V_{i_{\ell-1}}$ and set
\begin{equation*}
V =\bigcup_{1\leq j\leq \ell-1} V_{i_j}.
\end{equation*}
Thus,
\begin{equation}\label{eqn28}
\begin{split}
|V|=&\sum_{1\leq j\leq \ell-1\atop | \tilde{V}_{i_j} |<\delta-1} |V_{i_j}|+\sum_{1\leq j\leq \ell-1\atop | \tilde{V}_{i_j} |\geq \delta-1} |V_{i_j}|\\
=& T+\sum_{1\leq j\leq \ell-1\atop | \tilde{V}_{i_j} |\geq \delta-1} (\delta-1)\\
\geq & T+(\delta-1)\left\lfloor\frac{|\tilde{V}|-T}{r+\delta-1}\right\rfloor\\
\geq &(\delta-1)\left\lfloor\frac{|\tilde{V}|}{r+\delta-1}\right\rfloor,
\end{split}
\end{equation}
where $$T=\sum_{1\leq j\leq \ell-1\atop | \tilde{V}_{i_j} |<\delta-1} |V_{i_j}|,$$
 and the first inequality holds by the fact that in each round the set $\tilde{V}$ reduces
$|V_{i_j}|$ for the case $| \tilde{V}_{i_j} |<\delta-1$ and at most $|S_{i_j}|\leq r+\delta-1$ when $| \tilde{V}_{i_j} |\geq \delta-1$.

Denote $\cV_1 \triangleq \{j : |V_{i_j}|=\delta-1, j\in[\ell-1]\}$ and $\cV_2 \triangleq \{ j : |V_{i_j}|<\delta-1, j\in[\ell-1]\}$. On the one hand, if $|\cV_1| \ge \lfloor\frac{|\tilde{V}|}{r+\delta-1}\rfloor$, then set $B$ as the union of any $\lfloor\frac{|\tilde{V}|}{r+\delta-1}\rfloor$ sets in $\{V_{i_j}\}_{j\in \cV_1}$. On the other hand, if $|\cV_1| < \lfloor\frac{|\tilde{V}|}{r+\delta-1}\rfloor$, by \eqref{eqn28}, there exists $\hat{V}_{i_j} \subseteq V_{i_j}$ for $j\in\cV_2$  such that $B = (\cup_{j\in\cV_1} V_{i_j}) \cup (\cup_{j\in\cV_2} \hat{V}_{i_j})$ and $|B| = (\delta-1)\lfloor\frac{|\tilde{V}|}{r+\delta-1}\rfloor$.

Let $B_j = V_{i_j}$ if $j\in \cV_1$ and $B_j = \hat{V}_{i_j}$ if $j\in \cV_2$. Therefore, $B = \cup_{1\le t \le \tau} B_t$ and $\tau \le |\cV_1| + |\cV_2|$ since $\hat{V}_{i_j}$ could be an empty set for $j\in \cV_2$. The expression $B_{t_1} \cap S_{t_2} = \emptyset$ for all $1\leq t_2 < t_1 \leq \tau$ in the condition (i) comes from \eqref{eqn_new_B1} and  \eqref{eqn_new_B2}. The inequality $|B_t| \le \delta-1$ in the condition (ii) holds by  \eqref{eqn_B_1} and \eqref{eqn_B_tau}. The inequality \eqref{eqn30} holds by \eqref{eqn_B_1}-\eqref{eqn_new_B2}. Thus, we have $B_j$s for $j\in [\tau]$ are the desirable subsets of $B$, which completes the proof.
\end{IEEEproof}

Now, with the results above, we proceed with the proof of the following improved lower bound on the access cost.

\begin{theorem}\label{thm2}
Let $\bm{\mathcal{C}}$ be a $(t,1)_q$ LRCC with parameters $(n_I,k,*,*;n_F,tk,r,\delta)_a$ where $\delta\ge2$. Assume that the final code $\mathcal{C}_F$ is a linear code with $(r,\delta)_a$-locality and minimum distance $d$, and the optimal Hamming distance $d_{\opt}^{(q)}(n,k,r,\delta) = n-k+1-\phi(k,r,\delta)$. Then, the write access cost is
\begin{equation}\label{eqn17}
    \rho_w \ge
    t\left( d+(t-1)k-1+\phi((t-1)k,r,\delta) \right) - (t-1)n_F.
\end{equation}
Denote $\Delta=n_F - 2d - (t-1)k + 2 -\phi((t-1)k,r,\delta)$. Then, the read access cost is %$\Delta=n_F-2d-(t-1)k+2-(\lceil\frac{(t-1)k}{r}\rceil-1)(\delta-1)$. Then, the read access cost is
\begin{equation*}
    \rho_r \ge \begin{cases}
                 tk, & \mbox{if } \Delta\le 0,\\
                 t\left(k-\Delta + (\delta-1) \left\lfloor \frac{\Delta}{r+\delta-1} \right\rfloor\right), & \mbox{otherwise}.
               \end{cases}
\end{equation*}
Moreover, if $d>n_I-k+1$, then the read access is at least $tk$.
\end{theorem}
\begin{IEEEproof}
According to Lemma \ref{lem6}, the inequality \eqref{eqn17} follows directly from $\rho_w = n_F-\sum_{i=1}^{t}|\mathcal{U}_{i,1}|$. Again, by Lemma \ref{lem6},  we have
\begin{equation*}
    |\mathcal{U}_{i,1} \setminus \mathcal{D}_{i,1}| \le |\mathcal{U}_{i,1}|
    \le n_F - d - (t-1)k + 1 -\phi((t-1)k,r,\delta),
    %n_F - d - (t-1)k + 1 - \left(\left\lceil\frac{(t-1)k}{r}\right\rceil-1\right)(\delta-1),
\end{equation*}
which implies
\begin{equation*}
    \Delta_i = |\mathcal{U}_{i,1}\setminus \mathcal{D}_{i,1}|-d+1 \le n_F - 2d - (t-1)k + 2 -\phi((t-1)k,r,\delta) \triangleq \Delta,
    %n_F - 2d - (t-1)k + 2 - \left(\left\lceil\frac{(t-1)k}{r}\right\rceil-1\right)(\delta-1) \triangleq \Delta,
\end{equation*}
where $\Delta_i=|\mathcal{U}_{i,1}\setminus\mathcal{D}_{i,1}|-d+1$ is first defined in Lemma \ref{lem7}. Thus, $\Delta_i\le\Delta$ for every $i\in[t]$.
According to Lemma \ref{lem8}, let $\tilde{V}=\mathcal{U}_{i,1}\setminus\mathcal{D}_{i,1}$ then $|\tilde{V}|=\Delta_i+d-1$ by $\Delta_i=|\mathcal{U}_{i,1}\setminus\mathcal{D}_{i,1}|-d+1$. Then, we can find a $B$ satisfying conditions (i), (ii) and $|B| = (\delta-1)\lfloor\frac{\Delta_i}{r+\delta-1}\rfloor$.
 This is to say that
\begin{align*}
\Delta_i - |B|
&= \Delta_i - (\delta-1) \left\lfloor \frac{\Delta_i}{r+\delta-1} \right\rfloor \\
&= (\delta-1) \left( \frac{\Delta_i}{\delta-1} - \left\lfloor \frac{\frac{\Delta_i}{\delta-1}}{\frac{r+\delta-1}{\delta-1}} \right\rfloor \right) \\
&= (\delta-1) \left\lceil \frac{\left(\frac{r+\delta-1}{\delta-1}-1\right)\frac{\Delta_i}{\delta-1}}{\frac{r+\delta-1}{\delta-1}} \right\rceil \\
&\le (\delta-1) \left\lceil \frac{\left(\frac{r+\delta-1}{\delta-1}-1\right)\frac{\Delta}{\delta-1}}{\frac{r+\delta-1}{\delta-1}} \right\rceil \\
&= \Delta - (\delta-1) \left\lfloor \frac{\Delta}{r+\delta-1} \right\rfloor.
\end{align*}
Then, by Lemma \ref{lem7}, we have
\begin{equation*}
|\mathcal{D}_{i,1}| \ge \begin{cases}
            k, & \mbox{if } \Delta\le0, \\
            k-\Delta + (\delta-1) \left\lfloor \frac{\Delta}{r+\delta-1} \right\rfloor, & \mbox{otherwise},
          \end{cases}
\end{equation*}
for every $i\in[t]$. Therefore, $$\rho_r = \sum_{i=1}^{t}|\mathcal{D}_{i,1}|\ge \begin{cases}
                 tk, & \mbox{if } \Delta\le 0, \\
                 t\left(k-\Delta + (\delta-1) \left\lfloor \frac{\Delta}{r+\delta-1} \right\rfloor\right), & \mbox{otherwise}.
               \end{cases}$$

When $d>n_I-k+1$, the results follow directly from Lemma \ref{lem7}.
\end{IEEEproof}

\begin{definition}
An LRCC is called access-optimal if and only if the read and write access cost both achieve the lower bound given by Theorem \ref{thm2}.
\end{definition}

\begin{corollary}\label{thm4}
Suppose $d_{\opt}^{(q)}(n,k,r,\delta)= n-k+1-\left(\left\lceil\frac{k}{r}\right\rceil-1\right)(\delta-1)$ corresponding to Lemma \ref{lem5}.  Let $\bm{\mathcal{C}}$ be a $(t,1)_q$ LRCC with parameters $(n_I,k,*,*;n_F,tk,r,\delta)_a$ where $\delta\ge2$. Assume that the final code $\mathcal{C}_F$ is a linear code with $(r,\delta)_a$-locality and minimum distance $d$. Then, the write access cost is
\begin{equation}\label{eqn16}
    \rho_w \ge t\left( d+(t-1)k-1+\left(\left\lceil\frac{(t-1)k}{r}\right\rceil-1\right)(\delta-1) \right) - (t-1)n_F.
\end{equation}
Denote $\Delta=n_F-2d-(t-1)k+2-(\lceil\frac{(t-1)k}{r}\rceil-1)(\delta-1)$. Then, the read access cost is
\begin{equation}\label{eqn31}
    \rho_r \ge \begin{cases}
                 tk, & \mbox{if } \Delta\le 0, \\
                 t\left(k-\Delta + (\delta-1) \left\lfloor \frac{\Delta}{r+\delta-1} \right\rfloor\right), & \mbox{otherwise}.
               \end{cases}
\end{equation}
Moreover, if $d>n_I-k+1$, then the read access is at least $tk$.
\end{corollary}

\begin{figure}
  \centering
  \includegraphics[scale=0.5]{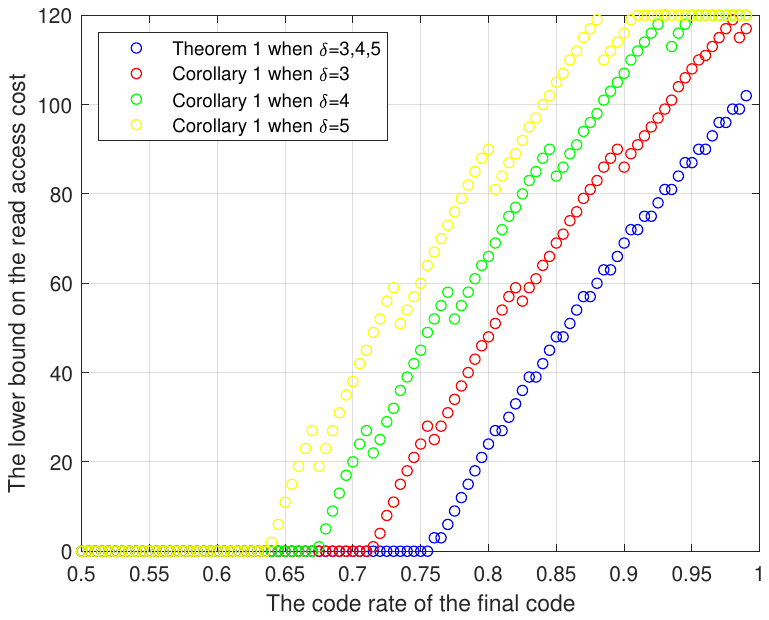}
  \caption{Compare the lower bounds on read access cost of Theorem \ref{thm1}~ \cite[Theorem IV.4]{kong2024locally} and Corollary \ref{thm4}. We consider only the case $\Delta>0$ and $d\le n_I-k+1$. Set the $t=3$ initial codes with parameters $[n_I=80,k=40]_q$, and the final code with parameters $[tk/\alpha,tk,d=15]_q$ and with $(r=10,\delta)_a$-locality where $\alpha$ is the code rate (the X-axis in the figure) of the final code. Note that the lower bound of Theorem \ref{thm1} is independent of the $\delta$, therefore, the conditions for $\delta=3,4,5$ are the same blue circle in the figure. }\label{fig_compare}
\end{figure}

\begin{remark}
Corollary \ref{thm4} extends the lower bound in Theorem \ref{thm1}~\cite[Theorem IV.4]{kong2024locally} which is only associated with $r$, to one related to $r$ and $\delta$.  For the case $\delta=2$, the new lower bound is exactly the same as the one in Theorem \ref{thm1}.
In the case $\delta>2$, the new lower bound on the write access cost clearly outperforms the one in Theorem \ref{thm1}. The expressions $\left\lfloor \frac{\Delta}{r+\delta-1} \right\rfloor, \ceilenv{ \frac{r\overline{\Delta}}{r+1} }$ are not easily reducible and LRCs achieving the minimum Hamming distance bound in Lemma \ref{lem5} usually require some parameter constraints \cite{song2014optimal,cai2022bound}, e.g., $r+\delta-1\mid n$, or $n=w(r+\delta-1)+m, k=ur+v$ such that $0<v<r, v+\delta-1\le m< r+\delta-1$, etc. Therefore, we investigate the following specific scenarios which show the new lower bound on the read access cost surpasses the one in Theorem \ref{thm1}, i.e., $\Delta - (\delta-1) \floorenv{\frac{\Delta}{r+\delta-1}} \le \ceilenv{ \frac{r\overline{\Delta}}{r+1} }$:
\begin{enumerate}
\item When $\Delta \mid (r+\delta-1)$, then
  $$ \Delta - (\delta-1) \floorenv{\frac{\Delta}{r+\delta-1}}
  = \frac{r\Delta}{r+\delta-1}
  < \frac{r\overline{\Delta}}{r+1}
  \le \ceilenv{ \frac{r\overline{\Delta}}{r+1} }, $$
  where the first inequality holds by $\Delta<\overline{\Delta}$ and $\delta>2$.
\item Let $\bm{\mathcal{C}}$ be with parameters $(n_I,kr,*,*;n_F,tkr,r,\delta)_a$ and the final code $\mathcal{C}_F$ be an optimal LRC satisfying $n_F=(tk+m)(r+\delta-1)$, where $m$ is a positive integer and $tk-m\ge 2$, then
    $$
      \Delta - (\delta-1) \floorenv{\frac{\Delta}{r+\delta-1}}
      = (k-m)r
      \le \ceilenv{ \frac{r\overline{\Delta}}{r+1} }
      = (k-m)r + \ceilenv{ \frac{ r((tk-2-m)(\delta-2)-1) }{r+1} }.
    $$
\end{enumerate}
The reader may refer to Figure \ref{fig_compare} as an example.
\end{remark}

\begin{remark}
We would like to comment that shortly after we published our results on arXiv \cite{ge2025Locally}, we learned that \cite{shi2025Bounds} has independently obtained a similar lower bound in Corollary \ref{thm4} focused on the initial codes with different parameters.
\end{remark}

In the following, we provide lower bounds for the access cost under two classes of parameters. These bounds correspond precisely to the parameter sets constructed in subsequent sections.

\begin{corollary}\label{cor7}
Suppose $d_{\opt}^{(q)}(n,k,r,\delta)= n-k+1-\left(\left\lceil\frac{k}{r}\right\rceil-1\right)(\delta-1)$ corresponding to Lemma \ref{lem5}. Let $\bm{\mathcal{C}}$ be a $(t,1)_q$ LRCC with parameters $((\ell/t+g-\ell)(r+\delta-1),\ell r/t,r,\delta; (\ell+h)(r+\delta-1), \ell r, r,\delta)_a$, where $t\mid\ell$, $\ell/t>g-\ell\ge h\ge0$ and $\delta\ge2$. Assume that the final code $\mathcal{C}_F$ is a linear code with $(r,\delta)_a$-locality and minimum distance $d=d_{\opt}^{(q)}$. Then the write access cost is at least $h(r+\delta-1)$ and the read access cost is at least $thr$.
\end{corollary}

\begin{proof}
Based on Corollary \ref{thm4}, the write access cost $\rho_w$ satisfies
\begin{align*}
\rho_w
&\ge t\left( d+(t-1)k-1+\left(\left\lceil\frac{(t-1)k}{r}\right\rceil-1\right)(\delta-1) \right) - (t-1)n_F \\
&= t(\delta+h(r+\delta-1)+(t-1)r\ell/t-1+((t-1)\ell/t-1)(\delta-1)) \\
&\quad -(t-1)(\ell+h)(r+\delta-1) \\
&= h(r+\delta-1),
\end{align*}
and
\begin{align*}
\Delta &= n_F-2d-(t-1)k+2- \left( \left\lceil \frac{(t-1)k}{r} \right\rceil-1 \right)(\delta-1) \\
&= (\ell+h)(r+\delta-1)-2(\delta+h(r+\delta-1)) \\
&\quad -(t-1)r\ell/t+2-((t-1)\ell/t-1)(\delta-1) \\
&= (\ell/t-h-1)(r+\delta-1) +r  \\
&> 0,
\end{align*}
where the last inequality holds by $\ell/t>g-\ell\ge h$. Therefore, the read access cost $\rho_r$ satisfies
\begin{align*}
\rho_r
&\ge t\left(r\ell/t-\Delta+(\delta-1) \left\lfloor\frac{\Delta}{r+\delta-1} \right\rfloor\right) \\
&= t(r\ell/t +h(r+\delta-1) -(r+\delta-1)\ell/t +\delta-1 + (\delta-1)(\ell/t-h-1)) \\
&= thr.
\end{align*}
\end{proof}

\begin{corollary}\label{cor6}
Suppose $d_{\opt}^{(q)}(n,k,r,\delta)= n-k+1-\left\lceil\frac{k}{r}\right\rceil(\delta-1)$ corresponding to Lemma \ref{lem1}. Let $\bm{\mathcal{C}}$ be a $(t,1)_q$ LRCC with parameters $((\ell/t+h)(r+\delta-1),\ell r/t,r,\delta;(\ell+h)(r+\delta-1),\ell r,r,\delta)_a$, where $\delta\ge2, t\mid \ell$ and $\ell/t>h\ge0$. Assume that the final code $\mathcal{C}_F$ is a linear code with $(r,\delta)_a$-locality and minimum distance $d=d_{\opt}^{(q)}$. Then the write access cost is at least $h(r+\delta-1)$ and the read access cost is at least $thr$.
\end{corollary}

\begin{proof}
By $\phi(k,r,\delta)=\left\lceil\frac{k}{r}\right\rceil(\delta-1)$ and Theorem \ref{thm2}, the write access cost $\rho_w$ satisfies
\begin{equation*}
\rho_w \ge n_F - \sum_{i\in[t]}|\cU_{i,1}|
= h(r+\delta-1)
\end{equation*}
and
\begin{align*}
\Delta &= n_F-2d-(t-1)k+2-\left\lceil\frac{(t-1)k}{r}\right\rceil(\delta-1) \\
&= \left(\frac{\ell}{t}-h\right)(r+\delta-1) \\
&>0.
\end{align*}
Therefore, the read access cost $\rho_r$ satisfies
\begin{align*}
\rho_r &\ge t\left(k-\Delta+(\delta-1) \left\lfloor\frac{\Delta}{r+\delta-1} \right\rfloor\right) \\
&= t\left( \frac{\ell r}{t} - \left(\frac{\ell}{t}-h\right)(r+\delta-1) + \left(\frac{\ell}{t}-h\right)(\delta-1) \right) \\
&= thr.
\end{align*}
\end{proof}

\begin{remark}
Corollaries \ref{cor7}, \ref{cor6} show the same lower bounds on access cost, although the value of $d_{\opt}^{(q)}$ is not the same. This inspires us to use the same method when constructing the access-optimal LRCCs with $d_{\opt}^{(q)}(n,k,r,\delta)= n-k+1-\left(\left\lceil\frac{k}{r}\right\rceil-1\right)(\delta-1)$ in Section \ref{sec_B} or $d_{\opt}^{(q)}(n,k,r,\delta)= n-k+1-\left\lceil\frac{k}{r}\right\rceil(\delta-1)$ in Section \ref{sec_A}.
\end{remark}

\section{A General Construction of LRCCs}\label{sec gen con}
In this section, we present a general construction of LRCCs with optimal access costs and optimal initial and final codes which achieve the minimum Hamming distance bound in \eqref{Sbound1}.
The basic idea of our construction is to choose the suitable initial and final codes from the punctured codes of a given base code.  To begin, we prove the following simple lemma.

\begin{lemma}\label{lem9}
Let $g,r,\delta,\ell$ be positive integers with $g\ge\ell$.
\begin{enumerate}
  \item For an optimal $[g(r+\delta-1),\ell r,d]_q$-LRC $\mathcal{C}$ with $(r,\delta)_a$-locality and achieving the minimum Hamming distance bound in \eqref{Sbound1}, the $\mathcal{C}$ admits a parity check matrix $H$ of the form \eqref{eqn20}, where $A_1,A_2,\dots,A_g$ are $(\delta-1)\times(r+\delta-1)$ matrices over $\mathbb{F}_q$, and $B_1,B_2,\dots,B_g$ are $((g-\ell) r)\times(r+\delta-1)$ matrices over $\mathbb{F}_q$.
  \item For any set $\mathcal{P}=\{i_1,\dots,i_{|\mathcal{P}|}\}\subseteq[g]$ with $|\mathcal{P}|>g-\ell$, the submatrix $H||_{\mathcal{P}}$ defined by \eqref{eqn19} is a parity check matrix for an optimal $[|\mathcal{P}|(r+\delta-1),(|\mathcal{P}|-g+\ell)r,d]_q$-LRC with $(r,\delta)_a$-locality and achieving the minimum Hamming distance bound in \eqref{Sbound1}.
\end{enumerate}
\end{lemma}
\begin{IEEEproof}
It directly follows from Lemma \ref{lem3} that $H$ is the parity check matrix of the code $\mathcal{C}$. Assume that a code $\tilde{\mathcal{C}}$ has the parity check matrix $H||_{\mathcal{P}}$. On the one hand, the minimum Hamming distance $d(\tilde{\mathcal{C}})$ of $\tilde{\mathcal{C}}$ satisfies $d(\tilde{\mathcal{C}})\ge d$, since any $d-1$ columns of $H||_\mathcal{P}$ are linearly independent over $\mathbb{F}_q$. On the other hand, the dimension $k(\tilde{\mathcal{C}})$ of $\tilde{\mathcal{C}}$ satisfies
\begin{align*}
k(\tilde{\mathcal{C}})
&= |\mathcal{P}|(r+\delta-1) - \rank(H||_{\mathcal{P}}) \\
&\ge |\mathcal{P}|(r+\delta-1) - (|\mathcal{P}|(\delta-1) + (g-\ell)r) \\
&= (|\mathcal{P}|-g+\ell)r,
\end{align*}
where the first inequality holds by the matrix $H||_{\mathcal{P}}$ has $|\mathcal{P}|(\delta-1)+(g-\ell)r$ rows. Note that $\tilde{\mathcal{C}}$ has $(r,\delta)_a$-locality. According to Singleton-type bound \eqref{Sbound1}, we have that
\begin{align*}
d(\tilde{\mathcal{C}})
&\le |\mathcal{P}|(r+\delta-1) - (|\mathcal{P}|-g+\ell)r +1 - (|\mathcal{P}|-g+\ell-1)(\delta-1) \\
&= g(r+\delta-1) - \ell r + 1 - (\ell-1)(\delta-1) \\
&= d,
\end{align*}
where the last equation holds by $\mathcal{C}$ is an optimal LRC. This implies that $d(\tilde{\mathcal{C}})=d$, i.e., $\tilde{\mathcal{C}}$ is an optimal LRC
with parameters $[|\mathcal{P}|(r+\delta-1),(|\mathcal{P}|-g+\ell)r,d]_q$.
\end{IEEEproof}

\begin{theorem}\label{thm3}
Let $g,\ell,r,\delta,t$ be positive integers such that $t\mid\ell$, $\ell/t>g-\ell\ge0$ and $\delta\ge2$. Suppose $\mathcal{C}$ is an optimal $[g(r+\delta-1),\ell r,d]_q$-LRC with $(r,\delta)_a$-locality. Then there exists an access-optimal LRCC with parameters
\begin{equation}\label{para1}
((\ell/t+g-\ell)(r+\delta-1),\ell r/t,r,\delta; (\ell+h)(r+\delta-1), \ell r, r,\delta)_a
\end{equation}
for any non-negative integer $h\le g-\ell$. Moreover, the initial and final codes achieve the minimum Hamming distance bound in \eqref{Sbound1}.
\end{theorem}
\begin{IEEEproof}
\textbf{Choose the initial codes}: Let $H$ be the parity check matrix of $\mathcal{C}$, where $H$ is of the form given in \eqref{eqn20}. Denote $s=\ell/t$. Let $\mathcal{P}_1,\dots,\mathcal{P}_t,\mathcal{P}$ be a partition of $[g]$ such that $|\mathcal{P}_i|=s$ for all $i\in[t]$ and $|\mathcal{P}|=g-\ell$. Without loss of generality, assume $\mathcal{P}_i=\{(i-1)s+1,\dots,is\}$ for $i\in[t]$ and $\mathcal{P}=\{\ell+1,\dots,g\}$. According to Lemma \ref{lem9} and $|\mathcal{P}_i\cup\mathcal{P}|=s+g-\ell$, the submatrix $H||_{\mathcal{P}_i\cup\mathcal{P}}$ defined by \eqref{eqn19} is a parity check matrix for an optimal $[(s+g-\ell)(r+\delta-1),sr,d]_q$-LRC $\mathcal{C}_i$ with $(r,\delta)_a$-locality for $i\in[t]$. Let $\mathcal{C}_i$ be the initial code with the parity check matrix $H||_{\mathcal{P}_i\cup\mathcal{P}}$ for $i\in[t]$.

\textbf{Puncture the initial codes}: Without loss of generality, assume $i=1$. We divide $H||_{\mathcal{P}_1\cup\mathcal{P}}$ into blocks as follows:
\begin{equation}\label{eqn26}
H||_{\mathcal{P}_1\cup\mathcal{P}} =
\left( \begin{array}{ccc|ccc|ccc}
  A_1 & \cdots & \bm{0} & \bm{0} & \cdots & \bm{0} & \bm{0} & \cdots & \bm{0} \\
  \vdots & \ddots & \vdots & \bm{0} & \cdots & \bm{0} & \bm{0} & \cdots & \bm{0} \\
  \bm{0} & \cdots & A_s & \bm{0} & \cdots & \bm{0} & \bm{0} & \cdots & \bm{0} \\
  \hline
  \bm{0} & \cdots & \bm{0} & A_{\ell+1} & \cdots & \bm{0} & \bm{0} & \cdots & \bm{0} \\
  \vdots & \ddots & \vdots & \vdots & \ddots & \vdots & \vdots & \ddots & \vdots \\
  \bm{0} & \cdots & \bm{0} & \bm{0} & \cdots & A_{\ell+h} & \bm{0} & \cdots & \bm{0} \\
  \hline
  \bm{0} & \cdots & \bm{0} & \bm{0} & \cdots & \bm{0} & A_{\ell+h+1} & \cdots & \bm{0} \\
  \vdots & \ddots & \vdots & \vdots & \ddots & \vdots & \vdots & \ddots & \vdots \\
  \bm{0} & \cdots & \bm{0} & \bm{0} & \cdots & \bm{0} & \bm{0} & \cdots & A_g \\
  B_{1,1} & \cdots & B_{s,1} & B_{\ell+1,1} & \cdots & B_{\ell+h,1} & B_{\ell+h+1,1} & \cdots & B_{g,1} \\
  \hline
  B_{1,2} & \cdots & B_{s,2} & B_{\ell+1,2} & \cdots & B_{\ell+h,2} & B_{\ell+h+1,2} & \cdots & B_{g,2}
\end{array} \right)
\triangleq\left(\begin{array}{c|c|c}
  W_{11} & \bm{0} & \bm{0} \\
  \hline
  \bm{0} & W_{12} & \bm{0} \\
  \hline
  M_{11} & M_{12} & M_{13} \\
  \hline
  N_{11} & N_{12} & N_{13}
\end{array}\right),
\end{equation}
where $B_{i,1}$ consists of the first $(g-\ell-h)r$ rows of $B_i$, and $B_{i,2}$ consists of the last $hr$ rows of $B_i$ for $i\in[g]$.

Since $\mathcal{C}_1$ is optimal with the minimum Hamming distance $d=\delta+(g-\ell)(r+\delta-1)$, therefore,
\begin{equation*}
\rank((M_{13}^\top, N_{13}^\top)^\top) = (g-\ell-h)(r+\delta-1).
\end{equation*}
There is an invertible square submatrix of $(M_{13}^\top, N_{13}^\top)^\top$ with order $(g-\ell-h)(r+\delta-1)$. Note that $M_{13}$ is a $(g-\ell-h)(r+\delta-1)\times (g-\ell-h)(r+\delta-1)$ matrix, and
$$ \rank\left(
\begin{pmatrix}
  A_{\ell+h+1} & \cdots & \bm{0} \\
  \vdots & \ddots & \vdots \\
  \bm{0} & \cdots & A_g
\end{pmatrix}
\right) = (g-\ell-h)(\delta-1). $$

By exchanging the last $(g-\ell)r$ rows of $H||_{\cP_1 \cup  \cP}$, the matrix $M_{13}$ can be invertible. We derive a punctured code $\mathcal{C}_1|_{[(s+h)(r+\delta-1)]}$ from $\mathcal{C}_1$ by performing elementary row operations on $H||_{\mathcal{P}_1\cup\mathcal{P}}$:
\begin{equation}\label{eqn22}
\begin{pmatrix}
    I & \bm{0} & \bm{0} & \bm{0}\\
    \bm{0} & I & \bm{0} & \bm{0} \\
    \bm{0} & \bm{0} & I & \bm{0} \\
    \bm{0} & \bm{0} & -N_{13}M_{13}^{-1} & I
\end{pmatrix}
\cdot
\begin{pmatrix}
  W_{11} & \bm{0} & \bm{0} \\
  \bm{0} & W_{12} & \bm{0} \\
  M_{11} & M_{12} & M_{13} \\
  N_{11} & N_{12} & N_{13}
\end{pmatrix}
=
\begin{pmatrix}
    W_{11} & \bm{0} & \bm{0} \\
    \bm{0} & W_{12} & \bm{0} \\
    M_{11} & M_{12} & M_{13} \\
    N_{11}-N_{13}M_{13}^{-1}M_{11} & N_{12}-N_{13}M_{13}^{-1}M_{12} & \bm{0}
 \end{pmatrix}
\triangleq H_1^*.
\end{equation}

In fact that the dual of puncturing code $\cC_1$ is equivalent to the shortening of the dual code $\cC_1^\bot$, and
\begin{equation*}
(\mathcal{C}_1|_{[(s+h)(r+\delta-1)]})^\bot
=\left\{
\bm{c}|_{[(s+h)(r+\delta-1)]} \middle|
\begin{array}{c}
  \bm{c}\in \bm{m} H_1^*,
  \bm{m}\in\F_q^{s(\delta-1)+(g-\ell)(r+\delta-1)}, \\
  \bm{c}|_{[(s+g-\ell)(r+\delta-1)]\setminus[(s+h)(r+\delta-1)]} = \bm{0}
\end{array}
\right\}.
\end{equation*}
Therefore, the punctured code $\mathcal{C}_1|_{[(s+h)(r+\delta-1)]}$ from \eqref{eqn22} has the parity check matrix
\begin{equation*}
    \overline{H}_1 = \begin{pmatrix}
        W_{11} & \bm{0} \\
        \bm{0} & W_{12} \\
        N_{11}-N_{13}M_{13}^{-1}M_{11} & N_{12}-N_{13}M_{13}^{-1}M_{12}
    \end{pmatrix}.
\end{equation*}

Since all the matrix $H||_{\mathcal{P}_i\cup\mathcal{P}}$ for $i\in[t]$ share the common submatrix $H||_{\mathcal{P}}$, thus the punctured code $\mathcal{C}_i|_{[(s+h)(r+\delta-1)]}$ from $\mathcal{C}_i$ comes from the same elementary row operations, i.e., each $H||_{\mathcal{P}_i\cup\mathcal{P}}$ is left-multiplied by the first matrix in the left-hand side of \eqref{eqn22}. Then, $\mathcal{C}_i|_{[(s+h)(r+\delta-1)]}$ has the parity check matrix $\overline{H}_i$ for $i\in[t]$, i.e.,
\begin{equation*}
    \overline{H}_i = \begin{pmatrix}
        W_{i1} & \bm{0} \\
        \bm{0} & W_{12} \\
        N_{i1}-N_{13}M_{13}^{-1}M_{i1} & N_{12}-N_{13}M_{13}^{-1}M_{12}
    \end{pmatrix},
\end{equation*}
which holds by $j_{12}=j_{22}=\cdots=j_{t2}$ for
$j\in\{W,M,N\}$, and $z_{13}=z_{23}=\cdots=z_{t3}$ for $z\in\{M,N\}$.

\textbf{Construct the final code}: To obtain the parity check matrix of the final code, we concatenate the parity check matrices $\overline{H}_1,\dots,\overline{H}_t$ and derive the following matrix:
\begin{equation}\label{eqn27}
H_F = \begin{pmatrix}
    W_{11} & \bm{0} & \cdots & \bm{0} & \bm{0} \\
    \bm{0} & W_{21} & \cdots & \bm{0} & \bm{0} \\
    \vdots & \vdots & \ddots & \vdots & \vdots \\
    \bm{0} & \bm{0} & \cdots & W_{t1} & \bm{0} \\
    \bm{0} & \bm{0} & \cdots & \bm{0} & W_{12} \\
    N_{11}-N_{13}M_{13}^{-1}M_{11} & N_{21}-N_{13}M_{13}^{-1}M_{21} & \cdots & N_{t1}-N_{13}M_{13}^{-1}M_{t1} & N_{12}-N_{13}M_{13}^{-1}M_{12}
\end{pmatrix}.
\end{equation}

Now, we construct the final code $\mathcal{C}_F$. Choose any $t$ codewords $\bm{c}_1\in\mathcal{C}_1,\dots,\bm{c}_t\in\mathcal{C}_t$, and calculate the new symbols as
\begin{equation}\label{eqn5}
\bm{c}^* = \sum_{i=1}^{t} \bm{c}_i|_{[(s+h)(r+\delta-1)]\setminus[s(r+\delta-1)]},
\end{equation}
where for $i\in[h]$,
\begin{equation}\label{eqn6}
A_{\ell+i} \cdot (\bm{c}^*|_{[i(r+\delta-1)]\setminus[(i-1)(r+\delta-1)]})^\top = \bm{0}
\end{equation}
and the code symbols $\bm{c}_i|_{[(s+h)(r+\delta-1)]\setminus[s(r+\delta-1)]}$ correspond to the read
symbols from $\bm{c}_i$ for $1\leq i\leq t$.
The final code $\mathcal{C}_F$ is defined as
\begin{equation*}
\mathcal{C}_F = \{ ( \bm{c}_1|_{[s(r+\delta-1)]}, \dots, \bm{c}_t|_{[s(r+\delta-1)]}, \bm{c}^* ) :  \bm{c}_1\in\mathcal{C}_1,\dots,\bm{c}_t\in\mathcal{C}_t \}.
\end{equation*}

\textbf{Verify that the final code is an optimal LRC}: Let the code $\tilde{\mathcal{C}}$ be given by the parity check matrix $H_F$. We first show that $\mathcal{C}_F\subseteq\tilde{\mathcal{C}}$. On the one hand, according to \eqref{eqn6} and the unchanged symbols satisfy the local parity check equations, for any codeword $\bm{c}\in\mathcal{C}_F$ and $i\in[\ell]$, we have
\begin{equation*}
( \underset{i-1 \text{ times}}{\underbrace{\bm{0},\cdots,\bm{0}}},A_i, \underset{\ell+h-i \text{ times}}{\underbrace{\bm{0},\cdots,\bm{0}}}) \cdot \bm{c}^\top = \bm{0}.
\end{equation*}
This implies that, for $i\in[t]$,
\begin{equation}\label{eqn7}
( \underset{i-1 \text{ times}}{\underbrace{\bm{0},\cdots,\bm{0}}},W_{i1}, \underset{t-i+1 \text{ times}}{\underbrace{\bm{0},\cdots,\bm{0}}}) \cdot \bm{c}^\top = \bm{0}.
\end{equation}
On the other hand,
\begin{equation}\label{eqn8}
\begin{split}
& \begin{pmatrix}
    \bm{0} & \cdots & \bm{0} & W_{12} \\
    N_{11}-N_{13}M_{13}^{-1}M_{11} & \cdots & N_{t1}-N_{13}M_{13}^{-1}M_{t1} & N_{12}-N_{13}M_{13}^{-1}M_{12}
\end{pmatrix}
\cdot \bm{c}^\top \\
&= \sum_{i=1}^{t} \begin{pmatrix} \bm{0} \\ N_{i1}-N_{13}M_{13}^{-1}M_{i1} \end{pmatrix} \cdot (\bm{c}_i|_{[s(r+\delta-1)]})^\top + \begin{pmatrix} W_{12} \\ N_{12}-N_{13}M_{13}^{-1}M_{12} \end{pmatrix} \cdot (\bm{c}^*)^\top \\
&= - \sum_{i=1}^{t} \begin{pmatrix} W_{12} \\ N_{12}-N_{13}M_{13}^{-1}M_{12} \end{pmatrix} \cdot (\bm{c}_i|_{[(s+h)(r+\delta-1)]\setminus[s(r+\delta-1)]})^\top + \begin{pmatrix} W_{12} \\ N_{12}-N_{13}M_{13}^{-1}M_{12} \end{pmatrix} \cdot (\bm{c}^*)^\top \\
&= \begin{pmatrix} W_{12} \\ N_{12}-N_{13}M_{13}^{-1}M_{12} \end{pmatrix} \cdot \left( \bm{c}^*-\sum_{i=1}^{t} \bm{c}_i|_{[(s+h)(r+\delta-1)]\setminus[s(r+\delta-1)]} \right)^\top \\
&= \bm{0},
\end{split}
\end{equation}
where the second equation holds by $\overline{H}_i \cdot (\bm{c}_i|_{[(s+h)(r+\delta-1)]})^\top = \bm{0}$ for $i\in[t]$, and the last equation comes from \eqref{eqn5}. Combining \eqref{eqn7} and \eqref{eqn8}, we have that $H_F \cdot \bm{c}^\top = \bm{0}$ for any $\bm{c}\in\mathcal{C}_F$. Therefore, $\mathcal{C}_F\subseteq\tilde{\mathcal{C}}$ and
\begin{equation}\label{eqn_dim_Cf}
\ell r=\dim(\mathcal{C}_F)\le\dim(\tilde{\mathcal{C}}).
\end{equation}

Secondly, we claim that $\mathcal{C}_F$ is an optimal $[(\ell+h)(r+\delta-1),\ell r,d]_q$-LRC.
By the fact $\mathcal{C}_F\subseteq\tilde{\mathcal{C}}$, it is sufficient to show that $\mathcal{C}_F=\tilde{\mathcal{C}}$ and $\tilde{\mathcal{C}}$ is an optimal LRC.
 Note that $H_F$ is also the parity check matrix of the punctured code $\mathcal{C}|_{[(\ell+h)(r+\delta-1)]}$ from $\mathcal{C}$ by \eqref{eqn22} and \eqref{eqn27}. Thus, $\tilde{\mathcal{C}}=\mathcal{C}|_{[(\ell+h)(r+\delta-1)]}$. Recall that $\mathcal{C}$ is an optimal LRC, which means
\begin{equation*}
d = g(r+\delta-1)-\ell r+1-\left( \left\lceil\frac{\ell r}{r}\right\rceil-1 \right)(\delta-1)
=(g-\ell)(r+\delta-1)+\delta,
\end{equation*}
by \eqref{Sbound1}.
Therefore, by  Lemma \ref{lem4}, we have $\dim(\tilde{\mathcal{C}})=\ell r$ since $(g-\ell-h)(r+\delta-1)<d$, which together with \eqref{eqn_dim_Cf} shows that $\mathcal{C}_F=\tilde{\mathcal{C}}$.

On the one hand, according to the puncturing operation on $\mathcal{C}$, the minimum Hamming distance of $\tilde{\mathcal{C}}$ satisfies
\begin{equation}\label{eqn23}
d(\tilde{\mathcal{C}}) \ge d-(g-\ell-h)(r+\delta-1) = \delta+h(r+\delta-1).
\end{equation}
On the other hand, $\tilde{\mathcal{C}}$ is an $[(\ell+h)(r+\delta-1),\ell r,d(\tilde{\mathcal{C}})]_a$-LRC with $(r,\delta)_a$-locality, by the Singleton-type bound \eqref{Sbound1}, we have
\begin{equation}\label{eqn24}
d(\tilde{\mathcal{C}})
\le (\ell+h)(r+\delta-1)-\ell r+1-\left(  \left\lceil \frac{\ell r}{r} \right\rceil-1 \right)(\delta-1) \\
= \delta + h(r+\delta-1).
\end{equation}
Combining \eqref{eqn23} and \eqref{eqn24}, the code $\tilde{\mathcal{C}}=\mathcal{C}_F$ is an optimal LRC.
Therefore, there exists an LRCC with parameters $((\ell/t+g-\ell)(r+\delta-1),\ell r/t,r,\delta; (\ell+h)(r+\delta-1), \ell r, r,\delta)_a$ as given in \eqref{para1}. The initial and final codes are all optimal LRCs.

\textbf{Verify that the access cost is optimal}: Combining \eqref{eqn5} and \eqref{eqn6}, we derive that the write access cost is $h(r+\delta-1)$ and the read access cost is $thr$. According to Corollary \ref{cor7}, we have that $\rho_w\ge h(r+\delta-1)$ and $\rho_r\ge thr$. Thus, the read and write access costs both achieve the lower bounds in Corollary \ref{cor7}. Thus, the LRCC above is access-optimal, which completes the
proof.
\end{IEEEproof}

In fact, the proof of Theorem \ref{thm3} gives the following general construction of access-optimal LRCCs.

\begin{construction}\label{con}
Let $\mathcal{C}$ be an optimal $[g(r+\delta-1),\ell r,d]_q$-LRC with $(r,\delta)_a$-locality, the parity check matrix $H$ of the from \eqref{eqn20}, where $\delta, g,\ell,r$ are positive integers with $\delta\ge 2$ and $ g-\ell\ge 0$. Let $t$ be a positive integer such that $t\mid\ell$, $\ell/t>g-\ell$, and $h$ be a non-negative integer with $h\le g-\ell$. The corresponding LRCC is given by the following steps:
\begin{itemize}
  \item First, generate initial optimal $[(\ell/t+g-\ell)(r+\delta-1), \ell r/t]_q$-LRC $\mathcal{C}_i$ by means of the parity check matrix $H||_{\mathcal{P}_i\cup\mathcal{P}}$ as \eqref{eqn26} for $i\in[t]$.
  \item Second, establish the conversion procedure function as \eqref{eqn5}.
  \item Finally, construct the final optimal $[(\ell+h)(r+\delta-1), \ell r]_q$-LRC $\mathcal{C}_F$ with $(r,\delta)_a$-locality by means of the parity check matrix $H_F$ as \eqref{eqn27}.
\end{itemize}
\end{construction}

\begin{remark}\label{rem4}
In general, the $t$ initial codes in Construction \ref{con} could be the distinct codes. In the following case, the initial codes may be the same one. When $h= |\mathcal{P}| = g-\ell$  and $H||_{\mathcal{P}_i} = D_i \cdot H||_{\mathcal{P}_1}$ for $i\in[t]$ where $D_i$ is an invertible diagonal matrix formed as
\begin{equation*}
D_i =
\begin{pmatrix}
  D_{i1} & \bm{0}_{s(\delta-1) \times hr} \\
  \bm{0}_{hr \times s(\delta-1)} & D_{i2}
\end{pmatrix}.
\end{equation*}
In this case, both $H||_{\mathcal{P}_i \cup \mathcal{P}} = A_i \cdot H||_{\mathcal{P}_1 \cup \mathcal{P}}$ and $H||_{\mathcal{P}_1 \cup \mathcal{P}}$ can be the parity check matrices of initial code $\mathcal{C}_i$, where
\begin{equation*}
A_i =
\begin{pmatrix}
  D_{i1} & \bm{0}_{s(\delta-1) \times h(\delta-1)} & \bm{0}_{s(\delta-1) \times hr} \\
  \bm{0}_{h(\delta-1) \times s(\delta-1)} & I_{h(\delta-1) \times h(\delta-1)} & \bm{0}_{h(\delta-1) \times hr} \\
  \bm{0}_{hr \times s(\delta-1)} & \bm{0}_{hr  \times h(\delta-1)} & D_{i2}
\end{pmatrix}.
\end{equation*}
Therefore, for every $1\le i\le t$,
\begin{equation*}
\overline{H}_i = H||_{\mathcal{P}_i \cup \mathcal{P}} =
\begin{pmatrix}
  W_{i1} & \bm{0} \\
  \bm{0} & W_{12} \\
  N_{i1} & N_{12}
\end{pmatrix}.
\end{equation*}
Since $\cC_1=\cC_2=\dots=\cC_t$ with parity check matrix $H||_{\mathcal{P}_1 \cup \mathcal{P}}$, then modify the conversion procedure function \eqref{eqn5} to
\begin{equation}\label{eqn4}
(\bm{c}^*)^\top =
\begin{pmatrix}
W_{12} \\ N_{12}
\end{pmatrix}^{-1}
\sum_{i=1}^{t} \left(
\begin{pmatrix} I_{h(\delta-1) \times h(\delta-1)} & \bm{0}_{h(\delta-1) \times hr} \\ \bm{0}_{hr \times s(\delta-1)} & D_{i2}
\end{pmatrix}
\begin{pmatrix}
W_{12} \\ N_{12}
\end{pmatrix}
\cdot (\bm{c}_i|_{[(s+h)(r+\delta-1)]\setminus[s(r+\delta-1)]})^\top \right),
\end{equation}
which shows \eqref{eqn7} and \eqref{eqn8} also hold.  Therefore, the new LRCC is access-optimal, and the initial and final codes are optimal LRCs, where the initial codes are equal.
\end{remark}

\section{Explicit Constructions of LRCCs}\label{sec exp con}
%Recall that the Corollary \ref{thm3-cor} gives the initial codes by means of the parity check matrix $H||_{\mathcal{P}_i\cup\mathcal{P}}$ as \eqref{eqn26} for $i\in[t]$, the final code by means of $H_F$ as \eqref{eqn27}, and the conversion procedure as \eqref{eqn5}. Also, the initial and final codes are all optimal LRCs, and the resultant LRCC is access-optimal.
In this section, we show two explicit constructions satisfying the final code preserves super-linear length or maximally repairable property when the initial codes have super-linear length or maximally repairable property respectively, based on Construction \ref{con} by employing some base code $\mathcal{C}$ such that (1) $\mathcal{C}$ is an optimal $[g(r+\delta-1),\ell r,d]_q$-LRC with $(r,\delta)_a$-locality, where $\delta\ge 2$ and $g-\ell\ge0$; (2) $\mathcal{C}$ has the parity check matrix $H$ of the from \eqref{eqn20}.

\subsection{Super-linear length}\label{sec_A}

For MDS codes, the MDS conjecture implies that the length of MDS codes may not exceed $q+1$ ($q+2$ for $2\mid q$ and $k=3$), which means that
for MDS convertible codes the length of the final codes may be upper bounded by $q+1$ ($q+2$ for $2\mid q$ and $k=3$).
Analogously, Kong in \cite{kong2024locally} raises an interesting question about how large the code length, under the fixed size of the finite field, can be to ensure the existence of LRCCs with optimal access costs.

The trivial answer to this question is the length of the final code for LRCCs cannot exceed the optimal length of LRC referring to the known bounds in \cite{cai2020optimal,kong2021new}.

\begin{lemma}[{\cite[Theorem 2]{cai2020optimal}}]\label{theorem_bound_delta>2}
Let $n=w(r+\delta-1)$, $\delta\geq 2$, $k=ur+v$, and additionally,
$r|k$ or $u\geq 2(r+1-v)$, where all parameters are integers. Assume there
exists an optimal $[n,k,d]_q$ linear code $\cC$ with all symbol
$(r,\delta)$-locality, and define $t=\floorenv{(d-1)/\delta}$.  If
$2t+1>4$, then
\begin{align*}
n&\leq
\begin{cases}
\frac{r+\delta-1}{r}\parenv{\frac{t-1}{2(q-1)}q^{\frac{2(w-u)r-2v-2}{t-1}}+1}, &\text{ if } t \text{ is odd},  \\
\frac{t(r+\delta-1)}{2r(q-1)}q^{\frac{2(w-u)r-2v}{t}}, &\text{ if } t \text{ is even},\\
\end{cases}
\end{align*}
where $w-u$ can also be rewritten as $w-u=\lfloor(d-1+v)/(r+\delta-1)\rfloor$.
\end{lemma}

In this subsection, we propose an explicit construction, which implies that the above bound is also tight asymptotically for LRCCs for some special cases where the final code has super-linear length when the initial codes have super-linear length.

\textbf{Base Code A} (\cite[Construction A]{kong2021new}):
Let $d\ge\delta+1$ and $n=m(r+\delta-1)$, where $d,\delta,r,m$ are positive integers. For $i\in[m]$, let $G_i=\{g_{i,1},g_{i,2},\dots,g_{i,r+\delta-1}\}$ be a $(r+\delta-1)$-subset of $\mathbb{F}_q$. Then, for each $i\in[m]$, we can construct a $(d-1)\times (r+\delta-1)$ Vandermonde matrix with generating set $G_i$ of the form $(U_i^\top,V_i^\top)^\top$, where
\begin{equation*}
U_i = \begin{pmatrix}
        1 & 1 & \cdots & 1 \\
        g_{i,1} & g_{i,2} & \cdots & g_{i,r+\delta-1} \\
        \vdots & \vdots & \ddots & \vdots \\
        g_{i,1}^{\delta-2} & g_{i,2}^{\delta-2} & \cdots & g_{i,r+\delta-1}^{\delta-2}
      \end{pmatrix},
V_i = \begin{pmatrix}
        g_{i,1}^{\delta-1} & g_{i,2}^{\delta-1} & \cdots & g_{i,r+\delta-1}^{\delta-1} \\
        g_{i,1}^{\delta} & g_{i,2}^{\delta} & \cdots & g_{i,r+\delta-1}^{\delta} \\
        \vdots & \vdots & \ddots & \vdots \\
        g_{i,1}^{d-2} & g_{i,2}^{d-2} & \cdots & g_{i,r+\delta-1}^{d-2}
      \end{pmatrix}.
\end{equation*}
Let the code $\mathcal{C}$ have the following parity check matrix:
\begin{equation}\label{H}
H = \begin{pmatrix}
      U_1 & \bm{0} & \cdots & \bm{0} \\
      \bm{0} & U_2 & \cdots & \bm{0} \\
      \vdots & \vdots & \ddots & \vdots \\
      \bm{0} & \bm{0} & \cdots & U_m \\
      V_1 & V_2 & \cdots & V_m
    \end{pmatrix}.
\end{equation}

The ensuing lemma shows sufficient conditions for Base Code A to be optimal.

\begin{lemma}[{\cite[Theorem III.4]{kong2021new}}]\label{lem2}
Let $r+\delta\ge d\ge 2\delta+1$. Suppose that for any subset $S\subseteq[m]$ with $2\le |S|\le \lfloor\frac{d-1}{\delta}\rfloor$, then
\begin{equation}\label{G}
\left|\bigcup_{i\in S}G_i\right| \ge \left(r+\frac{\delta}{2}-1\right)|S| + \frac{\delta}{2},
\end{equation}
then any $d-1$ columns of $H$ given by \eqref{H} are linearly independent. As a result, the base code $\mathcal{C}$  is an optimal $[n,k,d]_q$-LRC with $(r,\delta)_a$-locality.
\end{lemma}

\begin{remark}
For more optimal locally repairable codes, the reader may refer to \cite{huang2013pyramid,gopalan2012locality,rawat2013optimal,tamo2014family,cadambe2015bounds,kim2018locally,
li2019construction,xing2019construction,cai2020optimal,cai2021optimal_GPMDS,chen2020improved,hao2020bounds,sasidharan2015codes} and references therein.
\end{remark}

Recall that the access cost in Construction A also achieves the bound in Corollary \ref{cor6}. Therefore, we can derive the following lemma and theorem.

\begin{lemma}\label{cor4}
Let $m\ge v\ge1$, and $i_1<i_2<\cdots<i_v$ where $i_j\in[m]$ for $j\in[v]$. Then the code $\mathcal{C}$ given by the parity check matrix $H||_{\{i_1,i_2,\dots,i_v\}}$ is an optimal $[v(r+\delta-1),(v-1)r,r+\delta]_q$-LRC with $(r,\delta)_a$-locality, where $H$ is the parity check matrix of the code $\tilde{\mathcal{C}}$ from Base Code A satisfying $r+\delta=d\ge2\delta+1$ and condition \eqref{G}.
\end{lemma}
\begin{IEEEproof}
Since $G_i,i\in[m]$ satisfies condition \eqref{G}, then $G_j,j\in{\{i_1,i_2,\dots,i_v\}} \subseteq [m]$ satisfies condition \eqref{G}. Also, the $\overline{d}$ for $\mathcal{C}$ satisfies $\overline{d}=d=r+\delta\ge2\delta+1$. According to Lemma \ref{lem2} and Lemma \ref{lem1}, the dimension $k(\mathcal{C})$ of $\mathcal{C}$ with parameters $[v(r+\delta-1),k(\mathcal{C}),\overline{d}= r+\delta]_q$ satisfies
\begin{equation*}
r+\delta = v(r+\delta-1)-k(\mathcal{C})+1 -\left\lceil \frac{k(\mathcal{C})}{r} \right\rceil (\delta-1)
\end{equation*}
and
\begin{equation*}
k(\mathcal{C}) \ge v(r+\delta-1)-(\delta-1)v - (\overline{d}-\delta) = (v-1)r.
\end{equation*}
Therefore, $k(\mathcal{C})=(v-1)r$.
\end{IEEEproof}

\begin{theorem}\label{cor5}
Let $u$ be a positive integer. Let the code $\mathcal{C}$ be a Base Code A satisfying $r+\delta=d\ge2\delta+1$ and condition \eqref{G}. If $u\mid m-1$, then there is an access-optimal LRCC given by Construction \ref{con} with parameters
\begin{equation*}
( (u+1)(r+\delta-1),ur,r,\delta; m(r+\delta-1), (m-1)r,r,\delta )_a.
\end{equation*}
Moreover, the initial and final codes have super-linear length and both are optimal LRCs with respect to the bound in Lemma \ref{lem1}.
\end{theorem}
\begin{IEEEproof}
In fact, $\mathcal{P}_i=\{(i-1)u+1,\dots,iu\}$ for $i\in[\frac{m-1}{u}]$, and $\mathcal{P}=\{m\}$. Based on \eqref{eqn5} and \eqref{eqn6}, the read access cost is $\rho_r=\frac{(m-1)r}{u}$ and the write access cost is $\rho_w=r+\delta-1$. The LRCC given by Construction \ref{con} is access-optimal, which corresponds to the case $h=1$ and $t=\frac{m-1}{u}$ for Corollary \ref{cor6}. According to Lemma \ref{cor4}, the initial codes are optimal $[(u+1)(r+\delta-1),ur,r+\delta]_q$-LRC with $(r,\delta)_a$-locality.

It is well-known that the base code $\mathcal{C}$ can have super-linear length \cite{kong2021new}. Therefore, the code length $m(r+\delta-1)$ satisfies the following:
\begin{equation*}
m(r+\delta-1) =
\begin{cases}
  \Omega(q^{\delta}), & \mbox{if } d\ge 2\delta+1, \\
  \Omega(q^{1+\frac{\delta}{2}-o(1)}), & \mbox{if } 2\mid\delta, 3\delta+1\le d\le 4\delta, \\
  \Omega(q^{\frac{\delta+1}{2}}), & \mbox{if } 2\nmid\delta, 3\delta+1\le d\le 4\delta, \\
  \Omega(q^{\frac{\delta}{2}} (q\log q)^{\frac{1}{\lfloor\frac{d-1}{\delta}\rfloor-1}}), & \mbox{if } 2\mid\delta, d\ge3\delta+1, \\
  \Omega(q^{\frac{\delta}{2}} (q\log q)^{\frac{1}{2(\lfloor\frac{d-1}{\delta}\rfloor-1)}} ), & \mbox{if } 2\nmid\delta, 2\mid\lfloor\frac{d-1}{\delta}\rfloor, d\ge3\delta+1, \\
  \Omega(q^{\frac{\delta}{2}} (q\log q)^{\frac{1}{2(\lfloor\frac{d-1}{\delta}\rfloor-2)}} ), & \mbox{if } 2\nmid\delta, 2\nmid\lfloor\frac{d-1}{\delta}\rfloor, \lfloor\frac{d-1}{\delta}\rfloor>3, d\ge3\delta+1.
\end{cases}
\end{equation*}
Since $(u+1)(r+\delta-1)=\frac{u+1}{m}\cdot m(r+\delta-1)$ and $\frac{u+1}{m}$ is a constant. Therefore, the initial codes can have super-linear length.
\end{IEEEproof}

\begin{remark}
For the case $2\delta+1=d$, $r=\delta+1$, the parameters in Lemma \ref{theorem_bound_delta>2} can be figured out as
$w-u=u+1-u=1$, $t=2$ and
$$n\leq \frac{2\delta}{\delta+1}q^{\delta+1}/(q-1)=O(q^{\delta}).$$
Thus, the code length $\Omega(q^{\delta})$ given by Theorem \ref{cor5} is asymptotically optimal with respect to
the bound in Lemma \ref{theorem_bound_delta>2}.
\end{remark}

\begin{remark}\label{rem1}
Generally speaking, the $t$ initial codes in Theorem \ref{cor5} could be the distinct codes. In the following case, the initial codes may be the same one. When arbitrary two of $\cup_{j\in\mathcal{P}_i} G_j$ for $i\in[t]$ belong to different cosets of one multiplicative group of $\mathbb{F}_q^*$ in Theorem \ref{cor5}, i.e.,
\begin{equation*}
\beta_{i_1}^{-1} \cdot \bigcup_{j \in\mathcal{P}_{i_1}} G_{j} = \beta_{i_2}^{-1} \cdot \bigcup_{j \in\mathcal{P}_{i_2}} G_{j},
\end{equation*}
where $i_1\ne i_2, i_1,i_2\in [t]$, and $\beta_i$ is a coset leader of the coset containing $\cup_{j\in\mathcal{P}_i} G_j$. Namely, the parity check matrix $H_i$ of the initial code $\cC_i$ for $i\in[t]$ in Theorem \ref{cor5} is
{\small
\begin{align*}
H_i&=\begin{pmatrix}
  U_{i,1} & \cdots & \mathbf{0} \\
  \vdots & \ddots & \vdots \\
  \mathbf{0} & \cdots & U_{i,u+1} \\
  V_{i,1} & \cdots & V_{i,u+1}
\end{pmatrix}\\
&=
\begin{pmatrix}
1 & \cdots & 0 & \cdots & 0 & \cdots & 0 & 0 & \cdots & 0 \\
\vdots & \ddots & \vdots & & \vdots &  & \vdots & \vdots &  & \vdots \\
0 & \cdots & \beta_i^{\delta-2} & \cdots & 0 & \cdots & 0 & 0 & \cdots & 0 \\
\vdots & & \vdots & \ddots & \vdots &  & \vdots & \vdots &  & \vdots \\
0 & \cdots & 0 & \cdots & 1 & \cdots & 0  & 0 & \cdots & 0\\
\vdots &  & \vdots & & \vdots & \ddots & \vdots & \vdots &  & \vdots \\
0 & \cdots & 0 & \cdots & 0 & \cdots & \beta_i^{\delta-2} & 0 & \cdots & 0 \\
0 & \cdots & 0 & \cdots & 0 & \cdots & 0 & \beta_i^{\delta-1} & \cdots & 0\\
\vdots &  & \vdots & & \vdots &  & \vdots & \vdots & \ddots & \vdots  \\
0 & \cdots & 0 & \cdots & 0 & \cdots & 0 & 0 & \cdots & \beta_i^{d-2}
\end{pmatrix}
\begin{pmatrix}
1 & \cdots & 1 & \cdots & 0 & \cdots & 0\\
\vdots & \ddots  & \vdots & & \vdots & & \vdots\\
\alpha_{1,1}^{\delta-2} & \cdots & \alpha_{1,r+\delta-1}^{\delta-2} & \cdots & 0 & \cdots & 0 \\
\vdots & & \vdots & & \vdots & & \vdots \\
0 & \cdots & 0 & \cdots & 1 & \cdots & 1 \\
\vdots &  & \vdots & & \vdots & \ddots & \vdots\\
0 & \cdots & 0 & \cdots & \alpha_{u+1,1}^{\delta-2} & \cdots & \alpha_{u+1,r+\delta-1}^{\delta-2} \\
\alpha_{1,1}^{\delta-1} & \cdots & \alpha_{1,r+\delta-1}^{\delta-1} & \cdots & \alpha_{u+1,1}^{\delta-1} & \cdots & \alpha_{u+1,r+\delta-1}^{\delta-1} \\
\vdots & & \vdots & & \vdots & & \vdots \\
\alpha_{1,1}^{d-2} & \cdots & \alpha_{1,r+\delta-1}^{d-2} & \cdots & \alpha_{u+1,1}^{d-2} & \cdots & \alpha_{u+1,r+\delta-1}^{d-2} \\
\end{pmatrix},
\end{align*}}where $\cup_{j\in[u+1]}\{\alpha_{j,1},\dots,\alpha_{j,r+\delta-1}\}$ is a multiplicative group of $\F_q^*$ and $\beta_i$ is a coset leader for $i\in[t]$. Therefore, the $t$ initial codes can be the same code by Remark \ref{rem4}.  If the initial code has super-linear length $n_I$, then the length of the final code generated by Theorem \ref{cor5} is $\Theta_t(tn_I)$ which can be super-linear.
\end{remark}

\begin{example}\label{exa1}
We give an example for access-optimal $(8,1)_{7^2}$ LRCC with parameters $u=2,r=2,\delta=2,d=4,m=17$ in Theorem \ref{cor5}. Since $x^2-3$ is an irreducible polynomial over $\F_7$, let $\F_{7^2}=\F_7[x]/(x^2-3)=\F_7(\beta)$, where $\beta^2-3=0$. Then, $\{1,\beta\}$ is a basis of $\F_{7^2}$ over $\F_7$. Clearly, $\F_7^*$ is a multiplicative subgroup of $\F_{7^2}^*$. There are $8$ cosets of $\F_7^*$, i.e., $\F_7^*$ and $(i+\beta)\F_7^*, i=0,1,\dots,6$. Let $G_0=\{5,6,0\}, G_1=\{1,2,3\}, G_2=\{3,4,5\}$. Obviously, $G_0,G_1,G_2$ satisfy the relationship \eqref{G}. Therefore, the corresponding initial code $\cC_0$, based on Base Code A and $G_0,G_1,G_2$, has the code length $9$ which is larger than the size of $\F_7$. According to Remark \ref{rem1}, let the initial code $\cC_i$ be given by Base Code A and $(i-1+\beta)G_0,(i-1+\beta)G_1,(i-1+\beta)G_2$ for $i\in\{1,2,3,\dots,7\}$. Then, the final code $\cC_F$, based on Base Code A and $G_0,G_1,G_2,(i-1+\beta)G_1\cup (i-1+\beta)G_2, i=1,2,\dots,7$, has the code length $(2\times 8+1)\times 3=51>49$, since there are $2$ groups of samples from each of $8$ initial codes and there is one group of samples corresponding to the global parity check symbols. It is easy to check that $\cC_F$ satisfies the requirement \eqref{G}. Therefore, the lengths of initial and final codes are both beyond the field size, which can not constructed by Tamo-Barg codes \cite{tamo2014family}. Also, the initial codes are the same one, i.e., $\cC_0=\cC_1=\dots=\cC_7$.
\end{example}

\subsection{Maximally recoverable codes}\label{sec_B}
In general, the local parity checks make it possible for LRCs to recover some predetermined erasure patterns whose size exceeds the minimum Hamming distance, e.g., maximally recoverable codes. However, it remains an open question whether LRCCs retain this property. In this subsection, we apply explicit constructions to address this question. More specifically, we consider the case of maximally recoverable codes, which can recover from all information-theoretically recoverable erasure patterns.

\begin{definition}[\cite{gopi2020maximally,cai2020construction}]
Let $\mathcal{C}$ be an $[n,k,d]_q$ code with $(r,\delta)_a$-locality, and define $\mathcal{S}=\{S_i:i\in[n]\}$, where $S_i$ is an $(r,\delta)$-repair set for coordinate $i$. The code $\mathcal{C}$ is said to be a maximally recoverable (MR) code if $\mathcal{S}$ is a partition of $[n]$, and for any $R_i\subseteq S_i$ such that $|S_i\setminus R_i|=\delta-1$, the punctured code $\mathcal{C}|_{\cup_{1\le i\le n}R_i}$ is an MDS code. Specifically, if each $S_i\in\mathcal{S}$ is of size $|S_i|=r+\delta-1$, the code $\mathcal{C}$ is said to be an $(n,r,h,\delta,q)$-MR code, where $m=\frac{n}{r+\delta-1}$ and $h=mr-k$.
\end{definition}

We introduce the linearized Reed-Solomon codes \cite{martinez2018skew}, where the main construction of MR codes in \cite{cai2020construction} is based on it.

Let $\mathbb{F}_q\subseteq\mathbb{F}_{q_1}$. For any $\alpha\in\mathbb{F}_{q_1}$ and $i\in\mathbb{N}$, the $\mathbb{F}_q$-linear operator $\mathcal{D}_\alpha^i:\mathbb{F}_{q_1}\to\mathbb{F}_{q_1}$ is defined by
\begin{equation*}
\mathcal{D}_\alpha^i(\beta) = \beta^{q^i} \prod_{j=0}^{i-1} \alpha^{q^j}.
\end{equation*}
Let $\alpha\in\mathbb{F}_{q_1}$ and $\boldsymbol{\beta}=(\beta_1,\beta_2,\dots,\beta_h)\in\mathbb{F}_{q_1}^h$. For $i\in\mathbb{N}\cup\{0\}$ and $k,\ell\in\mathbb{N}$, where $\ell\le h$, define the matrices
\begin{equation*}
D(\alpha^i,\boldsymbol{\beta},k,\ell) =
\begin{pmatrix}
\beta_1 & \beta_2 & \cdots & \beta_\ell \\
\mathcal{D}_{\alpha^i}^1(\beta_1) & \mathcal{D}_{\alpha^i}^1(\beta_2) & \cdots & \mathcal{D}_{\alpha^i}^1(\beta_\ell) \\
\vdots & \vdots & & \vdots \\
\mathcal{D}_{\alpha^i}^{k-1}(\beta_1) & \mathcal{D}_{\alpha^i}^{k-1}(\beta_2) & \cdots & \mathcal{D}_{\alpha^i}^{k-1}(\beta_\ell)
\end{pmatrix}
\in\mathbb{F}_{q_1}^{k\times\ell}.
\end{equation*}

In what follows, we introduce a construction for $(n,r,h,\delta,q)$-MR codes \cite{cai2020construction}.

\textbf{Base Code B}(\cite[Construction A]{cai2020construction}): Let $\alpha_1,\alpha_2,\dots,\alpha_{r+\delta-1}$ be $r+\delta-1$ distinct elements in $\mathbb{F}_q^*$. Let
\begin{equation*}
P = \begin{pmatrix}
1 & 1 & \cdots & 1 \\
\alpha_1 & \alpha_2 & \cdots & \alpha_{r+\delta-1} \\
\vdots & \vdots & & \vdots \\
\alpha_1^{\delta-2} & \alpha_2^{\delta-2} & \cdots & \alpha_{r+\delta-1}^{\delta-2}
\end{pmatrix}
\in\mathbb{F}_q^{(\delta-1)\times(r+\delta-1)}.
\end{equation*}
Let $\gamma_1,\gamma_2,\dots,\gamma_h\in\mathbb{F}_q^h$ form a basis of $\mathbb{F}_{q^h}$ over $\mathbb{F}_q$. Let $\boldsymbol{\beta}=(\beta_1,\beta_2,\dots,\beta_{r+\delta-1})$ where $\beta_i=\sum_{j=1}^{h} \gamma_j \alpha_i^{\delta-2+j}$ for $i\in[r+\delta-1]$. For $m\in\mathbb{N}$, let $\mathcal{C}$ be the linear code with length $n$ over $\mathbb{F}_{q^h}$ given by the parity check matrix
\begin{equation}\label{eqn25}
H = \begin{pmatrix}
P & \bm{0} & \cdots & \bm{0} \\
\bm{0} & P & \cdots & \bm{0} \\
\vdots & \vdots & \ddots & \vdots \\
\bm{0} & \bm{0} & \cdots & P \\
D(\gamma^0,\boldsymbol{\beta},h,r+\delta-1) & D(\gamma^1,\boldsymbol{\beta},h,r+\delta-1) & \cdots & D(\gamma^{m-1},\boldsymbol{\beta},h,r+\delta-1)
\end{pmatrix},
\end{equation}
where $\gamma\in\mathbb{F}_{q^h}$ is a primitive element.

\begin{lemma}[{\cite[Theorem 2]{cai2020construction}}]\label{cor2}
Let $q\ge\max\{r+\delta,m+1\}$. Then the code $\mathcal{C}$ from Base Code B is an $(n=m(r+\delta-1),r,h,\delta,q^h)$-MR code with minimum Hamming distance $d=(\lfloor\frac{h}{r}\rfloor+1)(\delta-1)+h+1$.
\end{lemma}

Therefore, we can derive the following theorem.

%\begin{lemma}
%Let $q\ge\max\{r+\delta,m+1\}$, $m\ge v\ge\lceil\frac{h}{r}\rceil$, and $i_1<i_2<\cdots<i_v$ where $i_j\in[m]$ for $j\in[v]$. Then the code $\mathcal{C}$ given by the parity check matrix $H||_{\{i_1,i_2,\dots,i_v\}}$ is an $(n=v(r+\delta-1),r,h,\delta,q^h)$-MR code, where $H$ is given by \eqref{eqn25}.
%\end{lemma}

\begin{theorem}\label{cor3}
Let $u$ be a positive integer and $q\ge\max\{r+\delta,m+1\}$. Let the code $\mathcal{C}$ be a Base Code B satisfying $r\mid h$, i.e., the dimension of $\mathcal{C}$ is a positive integer $k=(m-\frac{h}{r})r$. If $u\mid m-\frac{h}{r}$, then there is an access-optimal LRCC given by Construction \ref{con} with parameters
\begin{equation*}
( (u+h/r)(r+\delta-1), ur, r, \delta; m(r+\delta-1), k, r, \delta)_a.
\end{equation*}
Moreover, the initial and final codes are MR codes.
\end{theorem}
\begin{IEEEproof}
Recall that $\mathcal{P}_i=\{(i-1)u+1,\dots,iu\}$ for $i\in[\frac{m-h/r}{u}]$, and $\mathcal{P}=\{m-h/r+1,\dots,m\}$. This implies that the initial code $\cC_i$ has parity check matrix $H||_{\cP_i\cup\cP}$ for $i\in[t]$ and the final code $\cC_F$ is by means of the parity check matrix $H$, where $H$ is given by \eqref{eqn25}. Based on Lemma \ref{cor2} and $r\mid h$, the minimum Hamming distance of $\cC$ is $d=\delta+\frac{h}{r}(r+\delta-1)$ which achieves the lower bound \eqref{Sbound1}.  According to Theorem \ref{thm3} where $g=m$ and $\ell=m-h/r$, the LRCC given by Construction \ref{con} is access-optimal. Noting that $H||_{\cP_i\cup\cP}$ for $i\in[t]$ also maintain the structure of Base Code B, therefore, by Lemma \ref{cor2}, the initial codes are $((u+h/r)(r+\delta-1),r,h,\delta,q^h)$-MR codes and the final code is an $(m(r+\delta-1),r,h,\delta,q^h)$-MR code.
\end{IEEEproof}

\begin{remark}\label{rem2}
In general, the $t$ initial codes in Theorem \ref{cor3} could be the distinct codes. In the following case, the initial codes may be the same one. When $H||_{\mathcal{P}_i} = D_i \cdot H||_{\mathcal{P}_1}$, where
\begin{equation*}
D_i =
\begin{pmatrix}
I_{u(\delta-1) \times u(\delta-1)} & \bm{0}_{u(\delta-1)\times 1} & \bm{0}_{u(\delta-1)\times 1} & \cdots & \bm{0}_{u(\delta-1)\times 1} \\
\bm{0}_{1 \times u(\delta-1)} & 1 & 0 & \cdots & 0 \\
\bm{0}_{1 \times u(\delta-1)} & 0 & \prod_{j=0}^{1-1} (\gamma^{i-1})^{q^j} & \cdots & 0 \\
\vdots & \vdots & \vdots & \ddots & \vdots \\
\bm{0}_{1 \times u(\delta-1)} & 0 & 0 & \cdots & \prod_{j=0}^{h-1} (\gamma^{i-1})^{q^j} \\
\end{pmatrix},
\end{equation*}
the $t$ initial codes can be the same code by Remark \ref{rem4} provided that $h=r$.
\end{remark}

\begin{example}
We give an example for access-optimal $(2,1)_{7^2}$ LRCC with parameters $u=2,h=2,r=2,\delta=2,m=5$ in Theorem \ref{cor3}. Let $\F_{7^2}$ be the finite field in Example \ref{exa1}. It is easy to check that $\gamma=1+\beta$ is a primitive element. Recall that $\{1,\beta\}$ is a basis of $\F_{7^2}$ over $\F_7$. Let $\alpha_1=1,\alpha_2=2,\alpha_3=3$ and $\beta_i=1+\alpha_i^2\beta$ for $i=1,2,3$. Based on Remark \ref{rem2}, let the initial code $\cC_1$ be with the parity check matrix
{\small
\begin{equation*}
H_1 = \begin{pmatrix}
        1 & 1 & 1 & 0  & 0  & 0  & 0  &  0 & 0 \\
         0 & 0  & 0  & 1 & 1 & 1 &  0 &  0 &  0 \\
        0  & 0  &  0 & 0  &  0 &  0 & 1 & 1 & 1 \\
        1+\beta & 1+2^2\beta & 1+3^2\beta & 1+\beta & 1+2^2\beta & 1+3^2\beta & 1+\beta & 1+2^2\beta & 1+3^2\beta \\
        (1+\beta)^{7} & (1+2^2\beta)^{7} & (1+3^2\beta)^{7} & (1+\beta)^{7}\gamma & (1+2^2\beta)^{7}\gamma & (1+3^2\beta)^{7}\gamma & (1+\beta)^{7}\gamma^2 & (1+2^2\beta)^{7}\gamma^2 & (1+3^2\beta)^{7}\gamma^2
      \end{pmatrix}
\end{equation*}
}
and the initial code $\cC_2$ be with the parity check matrix
{\footnotesize
\begin{equation*}
H_2 = \begin{pmatrix}
        1 & 1 & 1 & 0  &  0 & 0  &0   & 0  & 0 \\
         0 & 0  &  0 & 1 & 1 & 1 &  0 &  0 & 0  \\
         0 & 0  & 0  & 0  &  0 & 0  & 1 & 1 & 1 \\
        1+\beta & 1+2^2\beta & 1+3^2\beta & 1+\beta & 1+2^2\beta & 1+3^2\beta & 1+\beta & 1+2^2\beta & 1+3^2\beta \\
        (1+\beta)^{7}\gamma^2 & (1+2^2\beta)^{7}\gamma^2 & (1+3^2\beta)^{7}\gamma^2 & (1+\beta)^{7}\gamma^3 & (1+2^2\beta)^{7}\gamma^3 & (1+3^2\beta)^{7}\gamma^3 & (1+\beta)^{7}\gamma^4 & (1+2^2\beta)^{7}\gamma^4 & (1+3^2\beta)^{7}\gamma^4
      \end{pmatrix}.
\end{equation*}
}
Note that $\cC_1=\cC_2$ since
\begin{equation*}
H_2 = \begin{pmatrix}
        1 & 0 & 0 & 0 & 0 \\
        0 & 1 & 0 & 0 & 0 \\
        0 & 0 & 1 & 0 & 0 \\
        0 & 0 & 0 & 1 & 0 \\
        0 & 0 & 0 & 0 & \gamma^2
      \end{pmatrix} H_1.
\end{equation*}
The final code $\cC_F$ given by the conversion procedure function \eqref{eqn4} has the parity check matrix
\begin{equation*}
H = \left(\begin{array}{cccccc|c}
        1 & 1 & 1 &  0 & 0  &   0 & \bm{0}_{1\times 9} \\
        0  & 0  &  0 & 1 & 1 & 1  & \bm{0}_{1\times 9}  \\
        \hline
        \bm{0}_{3\times 1}  &  \bm{0}_{3\times 1} & \bm{0}_{3\times 1}  & \bm{0}_{3\times 1}  & \bm{0}_{3\times 1}  &  \bm{0}_{3\times 1} &  \\
        1+\beta & 1+2^2\beta & 1+3^2\beta & 1+\beta & 1+2^2\beta & 1+3^2\beta & H_2 \\
        (1+\beta)^{7} & (1+2^2\beta)^{7} & (1+3^2\beta)^{7} & (1+\beta)^{7}\gamma & (1+2^2\beta)^{7}\gamma & (1+3^2\beta)^{7}\gamma &
      \end{array}\right).
\end{equation*}
It is easy to verify that $\cC_F$ is an MR code.
\end{example}

\section{Conclusion}\label{conclusion}
This paper studies the code conversion of locally repairable codes in the merge regime. We improved the lower bound on the access cost which is relevant to $(r,\delta\ge 2)$-locality, broadening the previously known result only related to $r$. As a consequence, we proposed a general construction of access-optimal LRCCs. Then, based on the general construction, we showed two explicit constructions of access-optimal LRCCs such that the final code achieves super-linear length when the initial codes have super-linear length, or the final code has maximally recoverable property when the initial codes have maximally recoverable property, for the first time. Notably, our construction represents the first solution applicable to $\delta > 2$ scenarios, surpassing previous constructions limited to $\delta
 = 2$.

\bibliographystyle{IEEEtranS}
\bibliography{HanBib}

% Generated by IEEEtranS.bst, version: 1.14 (2015/08/26)
\begin{thebibliography}{10}
\providecommand{\url}[1]{#1}
\csname url@samestyle\endcsname
\providecommand{\newblock}{\relax}
\providecommand{\bibinfo}[2]{#2}
\providecommand{\BIBentrySTDinterwordspacing}{\spaceskip=0pt\relax}
\providecommand{\BIBentryALTinterwordstretchfactor}{4}
\providecommand{\BIBentryALTinterwordspacing}{\spaceskip=\fontdimen2\font plus
\BIBentryALTinterwordstretchfactor\fontdimen3\font minus
  \fontdimen4\font\relax}
\providecommand{\BIBforeignlanguage}[2]{{%
\expandafter\ifx\csname l@#1\endcsname\relax
\typeout{** WARNING: IEEEtranS.bst: No hyphenation pattern has been}%
\typeout{** loaded for the language `#1'. Using the pattern for}%
\typeout{** the default language instead.}%
\else
\language=\csname l@#1\endcsname
\fi
#2}}
\providecommand{\BIBdecl}{\relax}
\BIBdecl

\bibitem{dimakis2010network}
A.~G. Dimakis, P.~B. Godfrey, Y.~Wu, M.~J. Wainwright, and K.~Ramchandran,
  ``Network coding for distributed storage systems,'' \emph{IEEE transactions
  on information theory}, vol.~56, no.~9, pp. 4539--4551, 2010.

\bibitem{gopalan2012locality}
P.~Gopalan, C.~Huang, H.~Simitci, and S.~Yekhanin, ``On the locality of
  codeword symbols,'' \emph{IEEE Transactions on Information Theory}, vol.~58,
  no.~11, pp. 6925--6934, 2012.

\bibitem{guruswami2017repairing}
V.~Guruswami and M.~Wootters, ``Repairing {Reed-Solomon} codes,'' \emph{IEEE
  transactions on Information Theory}, vol.~63, no.~9, pp. 5684--5698, 2017.

\bibitem{tamo2018repair}
I.~Tamo, M.~Ye, and A.~Barg, ``The repair problem for {Reed-Solomon} codes:
  Optimal repair of single and multiple erasures with almost optimal node
  size,'' \emph{IEEE Transactions on Information Theory}, vol.~65, no.~5, pp.
  2673--2695, 2018.

\bibitem{ye2017explicit}
M.~Ye and A.~Barg, ``Explicit constructions of high-rate mds array codes with
  optimal repair bandwidth,'' \emph{IEEE Transactions on Information Theory},
  vol.~63, no.~4, pp. 2001--2014, 2017.

\bibitem{tamo2017optimal}
I.~Tamo, M.~Ye, and A.~Barg, ``Optimal repair of {Reed-Solomon} codes:
  Achieving the cut-set bound,'' in \emph{2017 IEEE 58th Annual Symposium on
  Foundations of Computer Science (FOCS)}.\hskip 1em plus 0.5em minus
  0.4em\relax IEEE, 2017, pp. 216--227.

\bibitem{li2018generic}
J.~Li, X.~Tang, and C.~Tian, ``A generic transformation to enable optimal
  repair in {MDS} codes for distributed storage systems,'' \emph{IEEE
  Transactions on Information Theory}, vol.~64, no.~9, pp. 6257--6267, 2018.

\bibitem{li2016optimal}
J.~Li and X.~Tang, ``Optimal exact repair strategy for the parity nodes of the
  $(k+ 2, k) $ zigzag code,'' \emph{IEEE Transactions on Information Theory},
  vol.~62, no.~9, pp. 4848--4856, 2016.

\bibitem{tamo2012zigzag}
I.~Tamo, Z.~Wang, and J.~Bruck, ``Zigzag codes: {MDS} array codes with optimal
  rebuilding,'' \emph{IEEE Transactions on Information Theory}, vol.~59, no.~3,
  pp. 1597--1616, 2012.

\bibitem{zeh2016bounds}
A.~Zeh and E.~Yaakobi, ``Bounds and constructions of codes with multiple
  localities,'' in \emph{2016 IEEE International Symposium on Information
  Theory (ISIT)}.\hskip 1em plus 0.5em minus 0.4em\relax IEEE, 2016, pp.
  640--644.

\bibitem{liu2023generic}
Y.~Liu, J.~Li, and X.~Tang, ``A generic transformation to enable optimal
  repair/access {MDS} array codes with multiple repair degrees,'' \emph{IEEE
  Transactions on Information Theory}, 2023.

\bibitem{hu2017optimal}
Y.~Hu, X.~Li, M.~Zhang, P.~P.~C. Lee, X.~Zhang, P.~Zhou, and D.~Feng, ``Optimal
  repair layering for erasure-coded data centers,'' \emph{ACM Transactions on
  Storage (TOS)}, vol.~13, pp. 1--24, 2017.

\bibitem{hou2019rack}
H.~Hou, P.~P. Lee, K.~W. Shum, and Y.~Hu, ``Rack-aware regenerating codes for
  data centers,'' \emph{IEEE Transactions on Information Theory}, vol.~65,
  no.~8, pp. 4730--4745, 2019.

\bibitem{chen2020explicit}
Z.~Chen and A.~Barg, ``Explicit constructions of msr codes for clustered
  distributed storage: The rack-aware storage model,'' \emph{IEEE Transactions
  on Information Theory}, vol.~66, no.~2, pp. 886--899, 2020.

\bibitem{wang2023rack}
J.~Wang, D.~Zheng, S.~Li, and X.~Tang, ``Rack-aware msr codes with error
  correction capability for multiple erasure tolerance,'' \emph{IEEE
  Transactions on Information Theory}, vol.~69, no.~10, pp. 6428--6442, 2023.

\bibitem{DBLP:journals/tit/LiWHY24}
G.~Li, N.~Wang, S.~Hu, and M.~Ye, ``{MSR} codes with linear field size and
  smallest sub-packetization for any number of helper nodes,'' \emph{{IEEE}
  Trans. Inf. Theory}, vol.~70, no.~11, pp. 7790--7806, 2024.

\bibitem{huang2013pyramid}
C.~Huang, M.~Chen, and J.~Li, ``Pyramid codes: Flexible schemes to trade space
  for access efficiency in reliable data storage systems,'' \emph{ACM
  Transactions on Storage (TOS)}, vol.~9, no.~1, pp. 1--28, 2013.

\bibitem{rawat2013optimal}
A.~S. Rawat, O.~O. Koyluoglu, N.~Silberstein, and S.~Vishwanath, ``Optimal
  locally repairable and secure codes for distributed storage systems,''
  \emph{IEEE Transactions on Information Theory}, vol.~60, no.~1, pp. 212--236,
  2014.

\bibitem{tamo2014family}
I.~Tamo and A.~Barg, ``A family of optimal locally recoverable codes,''
  \emph{IEEE Transactions on Information Theory}, vol.~60, no.~8, pp.
  4661--4676, 2014.

\bibitem{cadambe2015bounds}
V.~R. Cadambe and A.~Mazumdar, ``Bounds on the size of locally recoverable
  codes,'' \emph{IEEE transactions on information theory}, vol.~61, no.~11, pp.
  5787--5794, 2015.

\bibitem{kim2018locally}
G.~Kim and J.~Lee, ``Locally repairable codes with unequal local erasure
  correction,'' \emph{IEEE Transactions on Information Theory}, vol.~64,
  no.~11, pp. 7137--7152, 2018.

\bibitem{li2019construction}
X.~Li, L.~Ma, and C.~Xing, ``Construction of asymptotically good locally
  repairable codes via automorphism groups of function fields,'' \emph{IEEE
  Transactions on Information Theory}, vol.~65, no.~11, pp. 7087--7094, 2019.

\bibitem{xing2019construction}
C.~Xing and C.~Yuan, ``Construction of optimal locally recoverable codes and
  connection with hypergraph,'' in \emph{46th International Colloquium on
  Automata, Languages, and Programming (ICALP 2019)}.\hskip 1em plus 0.5em
  minus 0.4em\relax Schloss Dagstuhl-Leibniz-Zentrum fuer Informatik, 2019.

\bibitem{cai2020optimal}
H.~Cai, Y.~Miao, M.~Schwartz, and X.~Tang, ``On optimal locally repairable
  codes with super-linear length,'' \emph{IEEE Transactions on Information
  Theory}, vol.~66, no.~8, pp. 4853--4868, 2020.

\bibitem{cai2021optimal_GPMDS}
H.~Cai and M.~Schwartz, ``On optimal locally repairable codes and generalized
  sector-disk codes,'' \emph{IEEE Transactions on Information Theory}, vol.~67,
  no.~2, pp. 686--704, 2021.

\bibitem{chen2020improved}
B.~Chen, W.~Fang, S.-T. Xia, J.~Hao, and F.-W. Fu, ``Improved bounds and
  {S}ingleton-optimal constructions of locally repairable codes with minimum
  distance 5 and 6,'' \emph{IEEE Transactions on Information Theory}, vol.~67,
  no.~1, pp. 217--231, 2021.

\bibitem{hao2020bounds}
J.~Hao, S.-T. Xia, K.~W. Shum, B.~Chen, F.-W. Fu, and Y.~Yang, ``Bounds and
  constructions of locally repairable codes: parity-check matrix approach,''
  \emph{IEEE Transactions on Information Theory}, vol.~66, no.~12, pp.
  7465--7474, 2020.

\bibitem{sasidharan2015codes}
B.~Sasidharan, G.~K. Agarwal, and P.~V. Kumar, ``Codes with hierarchical
  locality,'' in \emph{2015 IEEE International Symposium on Information Theory
  (ISIT)}.\hskip 1em plus 0.5em minus 0.4em\relax IEEE, 2015, pp. 1257--1261.

\bibitem{rawat2016locality}
A.~S. Rawat, D.~S. Papailiopoulos, A.~G. Dimakis, and S.~Vishwanath, ``Locality
  and availability in distributed storage,'' \emph{IEEE Transactions on
  Information Theory}, vol.~62, no.~8, pp. 4481--4493, 2016.

\bibitem{TaBaFr16bounds}
I.~Tamo, A.~Barg, and A.~Frolov, ``Bounds on the parameters of locally
  recoverable codes,'' \emph{IEEE Transactions on Information Theory}, vol.~62,
  no.~6, pp. 3070--3083, 2016.

\bibitem{cai2018optimal}
H.~Cai, M.~Cheng, C.~Fan, and X.~Tang, ``Optimal locally repairable systematic
  codes based on packings,'' \emph{IEEE Transactions on Communications},
  vol.~67, no.~1, pp. 39--49, 2019.

\bibitem{cai2019optimal}
H.~Cai, Y.~Miao, M.~Schwartz, and X.~Tang, ``On optimal locally repairable
  codes with multiple disjoint repair sets,'' \emph{IEEE Transactions on
  Information Theory}, vol.~66, no.~4, pp. 2402--2416, 2020.

\bibitem{jin2019construction}
L.~Jin, L.~Ma, and C.~Xing, ``Construction of optimal locally repairable codes
  via automorphism groups of rational function fields,'' \emph{IEEE
  Transactions on Information Theory}, vol.~66, no.~1, pp. 210--221, 2020.

\bibitem{mazumdar2014update}
A.~Mazumdar, V.~Chandar, and G.~W. Wornell, ``Update-efficiency and local
  repairability limits for capacity approaching codes,'' \emph{IEEE Journal on
  Selected Areas in Communications}, vol.~32, no.~5, pp. 976--988, 2014.

\bibitem{li2015framework}
J.~Li, X.~Tang, and U.~Parampalli, ``A framework of constructions of minimal
  storage regenerating codes with the optimal access/update property,''
  \emph{IEEE Transactions on Information theory}, vol.~61, no.~4, pp.
  1920--1932, 2015.

\bibitem{chen2020enabling}
Z.~Chen, M.~Ye, and A.~Barg, ``Enabling optimal access and error correction for
  the repair of {Reed-Solomon} codes,'' \emph{arXiv preprint arXiv:2001.07189},
  2020.

\bibitem{kadekodi2019cluster}
S.~Kadekodi, K.~Rashmi, and G.~R. Ganger, ``Cluster storage systems gotta have
  $\{$HeART$\}$: improving storage efficiency by exploiting disk-reliability
  heterogeneity,'' in \emph{17th USENIX Conference on File and Storage
  Technologies (FAST 19)}, 2019, pp. 345--358.

\bibitem{maturana2022convertible}
F.~Maturana and K.~Rashmi, ``Convertible codes: Enabling efficient conversion
  of coded data in distributed storage,'' \emph{IEEE Transactions on
  Information Theory}, vol.~68, no.~7, pp. 4392--4407, 2022.

\bibitem{maturana2023bandwidth}
------, ``Bandwidth cost of code conversions in distributed storage:
  Fundamental limits and optimal constructions,'' \emph{IEEE Transactions on
  Information Theory}, vol.~69, no.~8, pp. 4993--5008, 2023.

\bibitem{maturana2023locally}
------, ``Locally repairable convertible codes: Erasure codes for efficient
  repair and conversion,'' in \emph{2023 IEEE International Symposium on
  Information Theory (ISIT)}.\hskip 1em plus 0.5em minus 0.4em\relax IEEE,
  2023, pp. 2033--2038.

\bibitem{maturana2020access}
F.~Maturana, V.~C. Mukka, and K.~Rashmi, ``Access-optimal linear mds
  convertible codes for all parameters,'' in \emph{2020 IEEE International
  Symposium on Information Theory (ISIT)}.\hskip 1em plus 0.5em minus
  0.4em\relax IEEE, 2020, pp. 577--582.

\bibitem{kong2024locally}
X.~Kong, ``Locally repairable convertible codes with optimal access costs,''
  \emph{IEEE Transactions on Information Theory}, vol.~70, no.~9, pp.
  6239--6257, 2024.

\bibitem{ge2024mds}
S.~Ge, H.~Cai, and X.~Tang, ``{MDS} generalized convertible code,'' \emph{arXiv
  preprint arXiv:2407.14304}, 2024.

\bibitem{huffman2010fundamentals}
W.~C. Huffman and V.~Pless, \emph{Fundamentals of error-correcting
  codes}.\hskip 1em plus 0.5em minus 0.4em\relax Cambridge university press,
  2010.

\bibitem{xia2015tale}
M.~Xia, M.~Saxena, M.~Blaum, and D.~A. Pease, ``A tale of two erasure codes in
  $\{$HDFS$\}$,'' in \emph{13th USENIX conference on file and storage
  technologies (FAST 15)}, 2015, pp. 213--226.

\bibitem{ball2012Large}
S.~Ball, ``On large subsets of a finite vector space in which every subset of
  basis size is a basis,'' \emph{J. Eur. Math. Soc.}, vol.~14, no.~3, pp.
  733--748, 2012.

\bibitem{guruswami2019How}
V.~Guruswami, C.~Xing, and C.~Yuan, ``How long can optimal locally repairable
  codes be?'' \emph{IEEE Transactions on Information Theory}, vol.~65, no.~6,
  pp. 3662--3670, 2019.

\bibitem{kong2021new}
X.~Kong, X.~Wang, and G.~Ge, ``New constructions of optimal locally repairable
  codes with super-linear length,'' \emph{IEEE Transactions on Information
  Theory}, vol.~67, no.~10, pp. 6491--6506, 2021.

\bibitem{prakash2012optimal}
N.~Prakash, G.~M. Kamath, V.~Lalitha, and P.~V. Kumar, ``Optimal linear codes
  with a local-error-correction property,'' in \emph{2012 IEEE International
  Symposium on Information Theory Proceedings}.\hskip 1em plus 0.5em minus
  0.4em\relax IEEE, 2012, pp. 2776--2780.

\bibitem{song2014optimal}
W.~Song, S.~H. Dau, C.~Yuen, and T.~J. Li, ``Optimal locally repairable linear
  codes,'' \emph{IEEE Journal on Selected Areas in Communications}, vol.~32,
  no.~5, pp. 1019--1036, 2014.

\bibitem{cai2022bound}
H.~Cai and M.~Schwartz, ``A bound on the minimal field size of lrcs, and cyclic
  mr codes that attain it,'' \emph{IEEE Transactions on Information Theory},
  vol.~69, no.~4, pp. 2240--2260, 2022.

\bibitem{ge2025Locally}
S.~Ge, H.~Cai, and X.~Tang, ``Locally repairable convertible codes: Improved
  lower bound and general construction,'' \emph{arXiv preprint
  arXiv:2504.06734}, 2025.

\bibitem{shi2025Bounds}
H.~Shi, W.~Fang, and Y.~Gao, ``Bounds and optimal constructions of generalized
  merge-convertible codes for code conversion into lrcs,'' \emph{arXiv preprint
  arXiv:2504.09580}, 2025.

\bibitem{gopi2020maximally}
S.~Gopi, V.~Guruswami, and S.~Yekhanin, ``Maximally recoverable lrcs: A field
  size lower bound and constructions for few heavy parities,'' \emph{IEEE
  Transactions on Information Theory}, vol.~66, no.~10, pp. 6066--6083, 2020.

\bibitem{cai2020construction}
H.~Cai, Y.~Miao, M.~Schwartz, and X.~Tang, ``A construction of maximally
  recoverable codes with order-optimal field size,'' \emph{arXiv preprint
  arXiv:2011.13606}, 2020.

\bibitem{martinez2018skew}
U.~Mart{\'\i}nez-Pe{\~n}as, ``Skew and linearized reed--solomon codes and
  maximum sum rank distance codes over any division ring,'' \emph{Journal of
  Algebra}, vol. 504, pp. 587--612, 2018.

\end{thebibliography}
\end{document}